# Genetic Complexity in a *Drosophila* Model of Diabetes-Associated Misfolded Human Proinsulin

**Soo-Young Park[*,§], Michael Z. Ludwig[†,§], Natalia A. Tamarina[*], Bin Z. He[†,**], Sarah H. Carl[†], Desiree A. Dickerson[†], Levi Barse[†], Bharath Arun[†], Calvin Williams[†], Cecelia M. Miles[†,§§], Louis H. Philipson[*], Donald F. Steiner[*], Graeme I. Bell[*,‡], Martin Kreitman[†,1]**

[*] Department of Medicine, The University of Chicago, Chicago, IL 60637
[†] Department of Ecology and Evolution, The University of Chicago, Chicago, IL 60637
[‡] Department of Human Genetics, The University of Chicago, Chicago, IL 60637
[**] Current Address: FAS Center for Systems Biology, Harvard University, 52 Oxford Street, Cambridge, MA 02138
[§§] Current Address: Augustana College, 2001 S. Summit Avenue, Sioux Falls, SD 57197

[§] SYP and MZL contributed equally to this study.

[1] Corresponding author:

Martin Kreitman

Mailing address: Department of Ecology and Evolution, The University of Chicago, 1101 E 57th Street, Chicago, IL 60637-1573.

Phone: +1 773 971 6075. Fax: +1 773 702 9740.

Email: martinkreitman@gmail.com.

**ABSTRACT**

Recent studies on the genetics of complex human disease have been successful in identifying associated variants, but very few statistically significant associations have been resolved to the level of a specific gene, mutation or molecular mechanism. Also left unresolved are questions about the nature of this variation — the total number of loci involved, the magnitude of their effects, and their frequency distribution. Here, we describe a complementary approach using *Drosophila melanogaster* to create a genetic model of a human disease — permanent neonatal diabetes mellitus — and present experimental results describing dimensions of this complexity. The approach involves the transgenic expression of a misfolded mutant of human preproinsulin, $hINS^{C96Y}$, which is a cause of permanent neonatal diabetes. When expressed in fly imaginal discs, $hINS^{C96Y}$ causes a reduction of adult structures, including the eye, wing and notum. Eye imaginal discs exhibit defects in both the structure and arrangement of ommatidia. In the wing, expression of $hINS^{C96Y}$ leads to ectopic expression of veins and mechano-sensory organs, indicating disruption of wild type signaling processes regulating cell fates. These readily measurable "disease" phenotypes are sensitive to temperature, gene dose and sex. Mutant (but not wild type) proinsulin expression in the eye imaginal disc induces IRE1-mediated Xbp1 alternative splicing, a signal for endoplasmic reticulum (ER) stress response activation, and produces global change in gene expression. Mutant hINS transgene tester strains, when crossed to stocks from the Drosophila Genetic Reference Panel (DGRP) produces $F_1$ adults with a continuous range of disease phenotypes and large broad-sense heritability. Surprisingly, the severity of mutant hINS-induced disease in the eye is not correlated with that in the notum in these crosses, nor with eye reduction phenotypes caused by the expression of two dominant eye mutants acting in two different eye development pathways, *Drop* (*Dr*) or *Lobe* (*L*) when crossed into the same genetic backgrounds. The tissue specificity of genetic variability for mutant hINS-induced disease has, therefore, its own distinct signature. The genetic dominance of disease-specific phenotypic variability in our model of misfolded human proinsulin makes this approach amenable to genome-wide association study (GWAS) in a simple $F_1$ screen of natural variation.

**INTRODUCTION**

Model organisms are widely employed in mechanistic studies of human Mendelian disease (BEDELL et al. 1997a; BEDELL et al. 1997b; CHINTAPALLI et al. 2007; LIESCHKE and CURRIE 2007; OCORR et al. 2007; PASSADOR-GURGEL et al. 2007; SCHLEGEL and STAINIER 2007; LESSING and BONINI 2009). They are likewise an important resource for investigating the genetic underpinnings of continuously varying quantitative traits (PALSSON and GIBSON 2004; TELONIS-SCOTT et al. 2005; WANG et al. 2005; DWORKIN and GIBSON 2006; WANG et al. 2006; BERGLAND et al. 2008; GIBSON and REED 2008; AYROLES et al. 2009; DWORKIN et al. 2009; GOERING et al. 2009; MACKAY et al. 2009; MACKAY 2010; MACKAY 2011). Numerous models of human disease have been established in the fly (reviewed in (PANDEY and NICHOLS 2011)), including transgenic models of diseases ranging from neurodegeneration / retinal degeneration (BILEN and BONINI 2005; RYOO et al. 2007; LESSING and BONINI 2009; YU and BONINI) to cancer (RUDRAPATNA et al. 2012). Success with genetic screens to identify suppressors and enhancers of disease when mutants are over-expressed in a developing tissue, such as the eye-antennal imaginal disc, suggested to us that it might be possible to generate a fly model of misfolded-insulin-associated diabetes.

A number of dominant mutations in human proinsulin have been identified in patients with permanent neonatal diabetes (STOY et al. 2007; STOY et al. 2010). One class of these involves mutations leading to an unpaired cysteine. The mutation of Cys-96 to Tyr — hINS$^{C96Y}$ — abolishes a disulfide bridge between the A- and B-chains of the polypeptide, causing proinsulin to misfold and accumulate in the endoplasmic reticulum (ER). Induction of the unfolded protein response (UPR), caused by ER stress, ultimately leads to pancreatic beta-cell death (OYADOMARI et al. 2002; HARTLEY et al. 2010). Mutant-insulin-induced diabetes may also be a model for the more common type 2 (adult onset) form of diabetes, where increased demand for insulin overwhelms the pathways regulating protein folding and trafficking. In this case, the accumulation of misfolded wild type proinsulin in the ER is hypothesized to trigger pathways that respond to loss of proteostatic control (OYADOMARI et al. 2002; SCHEUNER and KAUFMAN 2008).

Many signaling mechanisms regulating proteostasis — the dynamics of protein expression and turnover including folding, processing, transport, regulation and degradation — are conserved between fly and human (Geminard et al. 2009; Karpac and Jasper 2009; Haselton and Fridell 2010; Biteau et al. 2011). Misfolded alleles of rhodopsin, for example, cause age-related retinal degeneration in both species. In the fly model, overexpression of *ninaE* (a mutant allele of the fly ortholog of human rhodopsin-1) in the eye-antennal imaginal disc induces ER-stress-associated UPR and pro-apoptotic signaling, resulting in adult-onset eye degeneration (Ryoo et al. 2007; Kang and Ryoo 2009; Mendes et al. 2009; Kang et al. 2012). Strongly conserved signaling mechanisms in these pathways led us to reason that over-expression of mutant human preproinsulin (hINS[C96Y]) in the fly would likewise unleash UPR and cell death, thus recapitulating biological processes acting in the human form of the disease.

To test this prediction we created a transgenic model of permanent neonatal diabetes in the fly by expressing hINS[C96Y] under regulatory control of the UAS-Gal4 system. We drove hINS expression in larval/pupal imaginal discs, precursors of adult structures, and measured the loss of adult tissue, expected if the mutant activated cell death pathways. We also examined phenotypes in flies expressing wild type human preproinsulin (hINS[WT]) as a control. Here we describe phenotypic characteristics of this Mendelian model of disease, including sex-specific differences, dosage, environmental sensitivity, and reorganization of gene expression.

Type 2 or adult-onset diabetes is a complex disease whose genetic underpinnings have proved challenging to identify in human studies (Taneera et al. 2012). A rationale for introducing a mutant gene into the fly to create a Mendelian model of disease, as for example with unstable polyglutamine triplet repeat proteins (Bilen and Bonini 2007; Lessing and Bonini 2008; Yu and Bonini), is to create a genotype that is "sensitized" to the presence of disease modifiers. What if, rather than screening for these modifiers in forward genetic screens with laboratory-produced mutants, the same genetically sensitized flies are used to screen wild genetic backgrounds for naturally occurring modifiers of disease? With *Drosophila* having 20-40 times greater density of single nucleotide polymorphisms (SNPs) than human, and being genetically variable for most phenotypic traits, we expected this genetic screen to expose abundant genetic variation for the severity of disease phenotypes.

We examined dominant and partially-dominant genetic variation for disease severity by crossing a panel of inbred lines derived from a natural population sample (DGRP, (MACKAY *et al.* 2012)) to a tester stock carrying both the mutant insulin transgene and an eye imaginal disc-specific Gal4 expression driver on the same chromosome (GMR>>hINS$^{C96Y}$). Measuring the effects of natural modifiers in outcrossed flies avoids inbreeding effects in the isogenic lines and better mimics their heterozygosity in natural populations, especially low-frequency variants. Repeated measurements of genetically identical F$_1$ flies also reduces non-genetic variance components compared to individual measurements, increasing the power to detect genetic differences (MACKAY *et al.* 2009). By examining adult eye reduction in F$_1$ flies, we quantify disease phenotypes in different genetic backgrounds, and describe its distribution of effects in a natural population sample.

We then investigated biological properties of the naturally occurring genetic variation unleashed by our model of proteostatic disease. We first determine the correlation structure of hINS$^{C96Y}$-induced phenotypes in the adult eye and notum when hINS$^{C96Y}$ is expressed in their respective imaginal discs in a set of DGRP lines. We provide evidence for different alleles or loci modifying the disease in the two tissues, contrary to our expectation that the same alleles would be acting. This result led us to investigate genetic variation acting in eye-specific developmental pathways. We measured eye reduction in the same DGRP lines in crosses to two classical dominant eye mutants, *Drop* (*Dr*) and *Lobe* (*L*), and found that both are also uncorrelated with eye reduction induced by hINS$^{C96Y}$ expression. The presence of tissue- and disease-specific modifiers in our model of a human Mendelian disease, affirms the suitability of *Drosophila* as a model for investigating genetically complex forms of human disease.

# MATERIALS AND METHODS

## *Drosophila stocks*

The Drosophila stocks used in this study are described in Table 1.

**Table 1** *Drosophila* stocks

| Stock | Genotype | Reference | Comment |
|---|---|---|---|
| **hINS transgene** | | | |
| WT-24; WT-6 | P(UAS-hINS$^{WT}$) w1118 background | This study | Wild type human proinsulin; $2^{nd}$ chromosome insertion site |
| M-1; M-101 | P(UAS-hINS$^{C96Y}$) w1118 background | This study | Mutant human proinsulin; $2^{nd}$ chromosome insertion site |
| **Gal4 drivers** | | | |
| GMR-Gal4 | w$^*$; P{Gal4-ninaE.GMR}12 | #1104 (Bloomington) | Expresses in eye disc morphogenetic furrow |
| ap-Gal4 | ap-Gal4/CyO | #25686 (Bloomington) | Expresses in developing mesothorax (notum) |
| en-Gal4 | en-Gal4 ciBe/CyO Act5c-GFP | From R. Fehon | Expressing in ventral compartment of wing imaginal disc |
| dpp-Gal4 | dpp$^{blnk}$-Gal4, UAS-GFP$^{NLS}$/TM6B | From R. Fehon | Expresses between dorsal and ventral compartments of wing imaginal disc |
| **[Gal4 driver], [hINS]; Gal-4 driver, hINS same chromosome** | | | |
| GMR>>hINS$^{WT}$ GMR>>hINS$^{C96Y}$ | w1118; GMR-Gal4, UAS-hINS$^{WT\ or\ C96Y}$, UAS-GFP / CyO | This study | GMR-Gal4 driver recombined onto hINS-bearing chromosome (WT-24 or M-1) |
| ap>>hINS$^{WT}$ ap>>hINS$^{C96Y}$ | w1118; ap-Gal4, UAS-hINS$^{WT\ or\ C96Y}$/ CyO | This study | ap-Gal4 driver recombined onto hINS-bearing chromosome (WT-24 or M-1) |
| **Other stocks** | | | |
| *Drop (Dr)* | w1118; Dr1 /TM3, twist-GFP | From R. Fehon | Reduced eye size; acts through Jak/Stat pathway in ventral eye development |
| *Lobe (L)* | L(1) | #318 (Bloomington) | Muscle segment homeobox-1 transcription factor; induces apoptosis in developing eye |
| Scutoid (Sco) | w1118; CyO dfd-YFP /sna$^{Sco}$ | From R. Fehon | Missing bristles on notum |
| DGRP | Inbred wild lines | Bloomington | "Core 38 " used in these experiments |

***Crosses***

Flies were maintained on standard commercial medium at 25°C. Mutant and wild type hINS phenotypes, including adult and imaginal disc morphology and gene expression, were examined in $F_1$ flies produced by crosses between stocks carrying a hINS transgene (M-1 or WT-24) and a tissue-specific Gal4 driver (GMR-Gal4, ap-Gal4, en-Gal4 or dpp-Gal4). To examine hINS phenotypes in outcrossed genetic backgrounds, we crossed DGRP inbred stocks with a "tester" stock in which a Gal4 driver (GMR-Gal4 or ap-Gal4) was recombined onto a second chromosome carrying a hINS transgene (designated GMR>>hINS or ap>>hINS; Table 1). For each cross, and also crosses between DGRP stocks and *L* or *Dr*, 5 healthy virgin females from the tester stock were mated with 5-10 healthy males from each the DGRP stocks. Parents were transferred to fresh culture bottles every two days for 8 days. Phenotypes were measured separately in a minimum of 10 individuals for each sex. Eye measurements were made on 3-5 day old adults only. This particular trait, however, is stable in adults and has good replicability (Figure S1). The crosses between the tester stock and DGRP stocks were generally carried out in a single block to minimize experimental error.

***Transgene construction and P-element-mediated transformation***

Gal4/UAS system was used for ectopic gene expression of the wild type and mutant (C96Y) human preproinsulin (S<small>T</small> J<small>OHNSTON</small> 2002). Transgenic human preproinsulin wild type (hINS$^{WT}$) and mutant (hINS$^{C96Y}$) flies were generated by subcloning the human preproinsulin cDNA (B<small>ELL</small> *et al.* 1979) into the Drosophila transformation vector pUAST (https://dgrc.cgb.indiana.edu/vectors/). Transformation was carried out as described in (S<small>PRADLING</small> *et al.* 1995). Mapping crosses are described in (L<small>UDWIG</small> *et al.* 1993). For the UAS-hINS$^{WT}$ and UAS-hINS$^{C96Y}$ constructs, we generated 8 and 19 independently transformed stocks, respectively, each of which contained a single transposon insertion. For each of two constructs (WT and C96Y) at least one insertion in each of the three major chromosomes of *D. melanogaster* was generated to control for the influence of position effect on transgene expression.

*Immunohistochemistry*

Drosophila third instar wandering larvae of either sex were dissected in phosphate buffered saline (PBS). Isolated discs (approximately 5 pairs per sample) were placed in a glass tube with 4% paraformaldehyde in PBS and incubated for 30-40 min at room temp. Discs were then washed 3 times in PBS, 5 min each, and treated with 1% Triton X-100 in PBS for 30 min at room temperature. Discs were washed again 3 times by PBS/5 min each and treated with 5% Normal Donkey Serum (NDS) in PBS. Staining with a mixture of mouse anti-human C-peptide (Millipore; 1:200) and rat anti-ELAV (Developmental Studies Hybridoma Bank, University of Iowa, IA; 1:200) was performed in PBS with 1% NDS. Secondary antibodies were from Jackson ImmunoResearch. After staining, imaginal discs were removed with a glass pipette coated with NDS, placed in a drop of "*SlowFade* Gold with DAPI" antifade solution (Invitrogen) and covered with a glass coverslip. Staining was observed with a Leica SP2 laser scanning confocal microscope with 20x or 63x objectives.

*Transcriptional profiling*

Total RNA from 12 eye imaginal discs from each stock was isolated from wandering third instar larvae using MELT$^{TM}$ Total Nucleic Acid Isolation System (Ambion). The quality and quantity of each RNA sample was checked using a 2100 BioAnalyzer (Agilent) and Nanodrop 1000 (Thermo Scientific). Amplification of total RNA and synthesis of cDNA was carried out using the Ovation$^{TM}$ RNA Amplification System V2 (NuGen Technologies) from 100 ng of total RNA. The amplified cDNA was purified using Zymo-Spin II Column (Zymo Research Clean and Concentrator$^{TM}$-25, Zymo Research). 3.75 µg of fragmented and labeled single-stranded cDNA targets were generated by the FL-Ovation$^{TM}$ cDNA Biotin Module V2 (NuGen) and hybridized to each Affymetrix-GeneChip$^{®}$ Drosophila Genome 2.0 Array. Four microarrays were used to estimate transcript levels for each of five genotypes (2 males and 2 females each): control (GMR-Gal4) expressing the Gal4 activator protein only; hINS$^{WT}$ line 6 (hereafter WT-6); hINS$^{WT}$ line 24 (WT-24); hINS$^{C96Y}$ line 101 (M-101); and hINS$^{C96Y}$ line 1 (M-1). The two lines of each genotype, hINS$^{WT}$ or hINS$^{C96Y}$, were selected to represent moderate and high expression of the hINS transgene. Microarray data are available at the Gene Expression Omnibus (GEO) (**http://www.ncbi.nlm.nih.gov/geo**) with accession number GSE43128.

### Analysis of microarray data

Intensity data for each feature on the array was calculated from the images generated by the GenChip® Scanner 3000 7G (Affymetrix) and the data files were extracted using GeneChip Operating Software (MicroArray Suite 5.0 software, Affymetrix). We performed background subtraction and normalization of CEL files both in dChip (2010.1) software with its default parameter (5th percentile of PM-probes as baseline for background subtraction, invariant set for normalization).

*Data analysis 1*: We used Partek software (Partek Inc., v6.5) initially to identify differentially expressed genes in the comparison between GMR-Gal4 background and each transgenic fly (GMR-Gal4/UAS-hINS$^{WT}$ or INS$^{C96Y}$, 4 genotypes). Sexes were analyzed separately. A one-way ANOVA was performed with genotype as a fixed effect. All genes for which the effect of genotype was significant at a false discovery rate (FDR) of 10% were further tested to determine whether mean expression of GMR-Gal4/UAS-hINS$^{WT}$ or GMR-Gal4/UAS-hINS$^{C96Y}$ was significantly different from the control (GMR-Gal4).

*Data* analysis 2: The normalized intensity data were log2 transformed for subsequent analyses implemented using R-bioconductor (v 2.10.0). To compare the transcriptional responses to the expression of wild type or mutant hINS, we restricted the analysis to a single pair of the transgenic lines, WT-24 and M-1, matched for high level of hINS expression based on quantitative real-time PCR (qRT-PCR), together with the control (GMR-Gal4), and performed independent ANOVA for each array feature under the model $Y_{ijk} = u + L_i + S_j + e_{ijk}$, where $L_i$ = line (i = 1, 2 or 3), $S_j$ = sex (j = 1 or 2) and $e_{ijk}$ = error (k = 1 or 2). We applied Benjamini–Hochberg–Yekutieli procedure (BENJAMINI AND YEKUTIELI 2001) on the resulting p-values to control FDR.

### Quantitative RT-PCR

Total RNA was isolated from heads of 30 adult flies using Trizol reagent (Invitrogen) and from eye-antennal imaginal discs from 12 wandering late-third instar larvae using

MELT^TM Total Nucleic Acid Isolation System (Ambion). cDNA synthesis was performed using oligo dT-primer and Superscript® III First-Strand Synthesis System (Invitrogen). Quantitative RT-PCR was carried out using a StepOne™Real-Time PCR System (ABI) in triplicate. Gene-specific sets of primers (Table S1) and SYBR green PCR master mix (ABI) were used to quantify gene expression. The results was normalized to the expression of *rp49* expression.

### Analysis of Xbp1 mRNA splicing

Total RNA was isolated from wandering late-third instar larvae and cDNA synthesized, as described above. To visualize the alternative splicing of the 23 bp Xbp1 intron, a diagnostic marker of UPR induction (Cox and Walter 1996; Mori *et al.* 2000; Shen *et al.* 2001; Yoshida *et al.* 2001), PCR was carried out using the *D. melanogaster*-specific primers 5'- AACAGCAGCACAACACCAGA-3' (forward) and 5'-CGCCAAGCATGTCTTGTAGA-3' (reverse), which amplifies fragments of 239 bp for unspliced Xbp1 (Xbp1-U) and 216 bp for spliced Xbp1 (Xbp1-S). The PCR conditions were initial denaturation at 94°C for 3 min, and 35 cycles of denaturation at 94°C for 30 sec, annealing at 57°C for 30 sec and extension at 72°C for 1 min, and a final extension at 72°C for 10 min. PCR products were separated by 10% PAGE and visualized by ethidium bromide staining. PCR products were also digested with *Pst*I to better distinguish the spliced and unspliced forms of Xbp1 mRNA variants.

### Eye measurement

Low magnification images were captured with a Zeiss AxioCam HRc mounted on a Leica MZ16 fluorescent stereomicroscope. For high magnification images eyes were mounted in Halocarbon 700 oil (Sigma) and were captured with a Zeiss AxioCam HRc camera on the Zeiss Axioscope microscope.

Three to five day old adults of the appropriate genotype from each cross (in some experiments following thorax measurement) were placed on a slide containing a thin layer of silicone vacuum grease (Beckman), and mounted in Halocarbon 700 oil under a cover slip supported by capillary tubes. Eyes were photographed using a Leica DFC420 camera mounted on a Leica M205 FA stereomicroscope. The Leica Application Suite software and ImageJ software (rsb.info.hih.gov/ij/) were used to analyze merged Z-stacks taken on the Leica M205 FA microscope. Only eyes with borders and head capsules in the same optical section were analyzed. At least 10 females and (or) males from a given cross were measured to obtain the average for each genotype. Eye area measurements are robust over across independent experiments (Figure S1).

### Thorax measurement

Thorax lengths (the distance from the base of the most anterior humeral bristle to the posterior tip of the scutellum) were measured using a Nikon SMZ-2B microscope equipped with a mechanical stage and built-in micrometer.

### Wing measurement

For each cross 5 healthy virgin females from both the dpp-Gal4 and en-Gal4 stocks were mated with 5-10 healthy males from w1118 (control), WT-24 and M-1 stocks. Flies of the appropriate genotype were incubated in 70% ethanol for at least 24 hr. Wings were mounted in Aqua PolyMount (Polysciences, Inc.) on a glass slide. Only wings that had been flattened during the mounting process were used for further analysis.  Images were captured using a Zeiss Axioscop microscope equipped with an AxioCam HRc camera and imported into Image J for analysis. Wing sizes in case of the en-Gal4 driver were quantified by measuring the posterior area divided by total wing area. Wing sizes in case of the dpp-Gal4 driver were quantified by measuring the area of sectors L4 and L3 area divided by sectors L2-L4 area. Significance was determined by the Mann-Whitney U test.

### Bristle count

Three to five days old flies of the appropriate genotype (25 males and 25 females) were placed in Halocarbon 700 oil. The presence or absence of 26 bristles (macrochaetae; Figure S2) on the notum, including humeri, was scored.

## RESULTS

### *Transgene analysis*

We generated transgenic flies carrying a single copy of either wild type or mutant human preproinsulin (hINS$^{WT}$ and hINS$^{C96Y}$, respectively), whose expression is regulated by a UAS:Gal4 promoter. A total of 27 independent transgenic stocks were produced, eight carrying hINS$^{WT}$ and 19 carrying hINS$^{C96Y}$, allowing us to investigate and control for position effects on gene expression. The hINS$^{WT}$ lines also gave us the ability to identify mutation-dependent phenotypes in hINS$^{C96Y}$ distinct from those resulting from both protein over-expression and/or interactions with native *Drosophila* Insulin Like Peptides (DILPs) dependent pathways.

We investigated disease phenotypes by expressing transgenic hINS in imaginal discs of the eye (GMR-Gal4 driver), wing (en-Gal4 and dpp-Gal4 drivers), and notum (ap-Gal4 driver). The eye system was studied in greater detail than the others. GMR-Gal4 directs hINS transgene expression to developing photoreceptor neurons and surrounding support eye cells in the eye-morphogenetic furrow (http://flystocks.bio.indiana.edu/Reports/9146.html) (FREEMAN 1996). We confirmed transgene expression in F$_1$ adult heads by qRT-PCR (data not shown) and the presence of hINS protein in late 3$^{rd}$ instar larval imaginal discs of GMR-Gal4 / UAS-hINS$^{C96Y}$ (or UAS-hINS$^{WT}$) individuals by immunoflourescent staining with an antibody specific for hINS C-peptide (Figure 1 I-L).

We then examined the adult eye phenotypes caused by the mutant transgene expression in comparison to GMR-Gal4 controls. All adult GMR-Gal4 and GMR-Gal4/UAS-hINS$^{WT}$ (eight lines) flies exhibited completely wild type eyes. In contrast, five of the mutant GMR-Gal4/UAS-hINS$^{C96Y}$ lines showed a severe eye phenotype: a reduction in eye size, reduced number of eye bristles, the presence of lesions with no evidence of cells, and collapse of ommatidia structure and normal array pattern. Six lines showed a moderate eye phenotype with a small reduction in eye size and eight lines had no obvious phenotype.

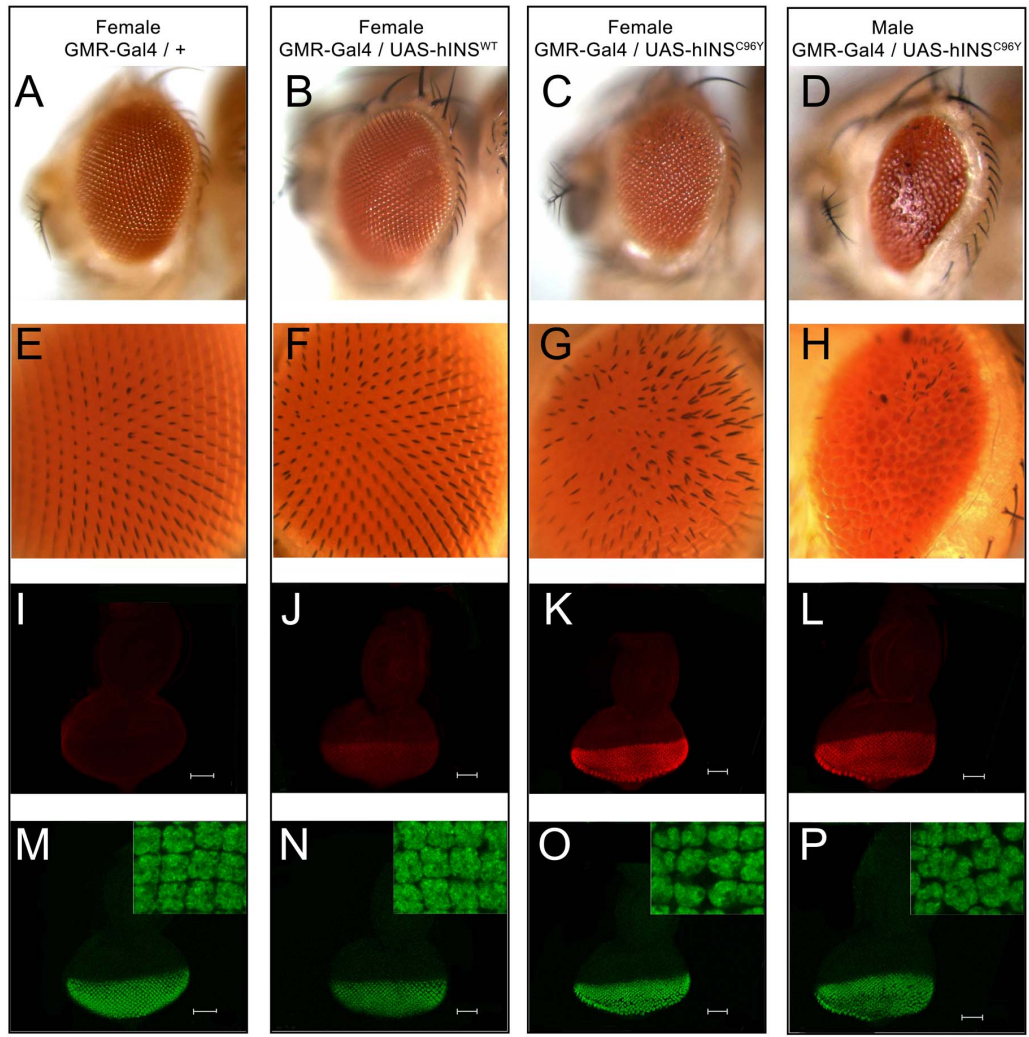

**Figure 1** Eye phenotypes induced by hINS$^{C96Y}$ transgene expression. **(A-D)** Eyes of 3-5 day-old adults. (A) Female, GMR-Gal4. (B) Female, GMR-Gal4 / UAS-hINS$^{WT}$. (C) Female, GMR-Gal4 / UAS-hINS$^{C96Y}$. (D) Male, GMR-Gal4 / UAS-hINS$^{C96Y}$. (E-H) High magnification images of adult eyes above showing defects in patterning of ommatidia and mechanosensory bristles. (I-H) Eye-antennal imaginal discs of third-instar larvae of genotypes noted in (A-D) stained with anti-human C-peptide antibody (red). (M-P) Discs in (panels I-H above) stained with anti-ELAV antibody (green); insets (panels M-P) show enlarged area of most posterior part of eye disc.

We first tested and could reject the formal possibility that the mutant insulin transgenes are expressed at higher levels than the wild type hINS transgenes, as could be caused by insertion-site position effects. hINS transcript levels were quantified by qRT-PCR from total RNA of late 3$^{rd}$ instar larval eye-imaginal discs from two mutant hINS lines, one exhibiting a mild eye phenotype (M-101) and the other a more severe eye phenotype (M-1), and found significantly greater expression in the line with the more severe eye phenotype (Figure 2A). This comparison establishes the C96Y mutation as being both necessary and sufficient to cause an eye degeneration phenotype. Subsequent analyses were carried out with a high-expressing mutant hINS line (M-1) and a matched wild type-hINS-expressing control (WT-24).

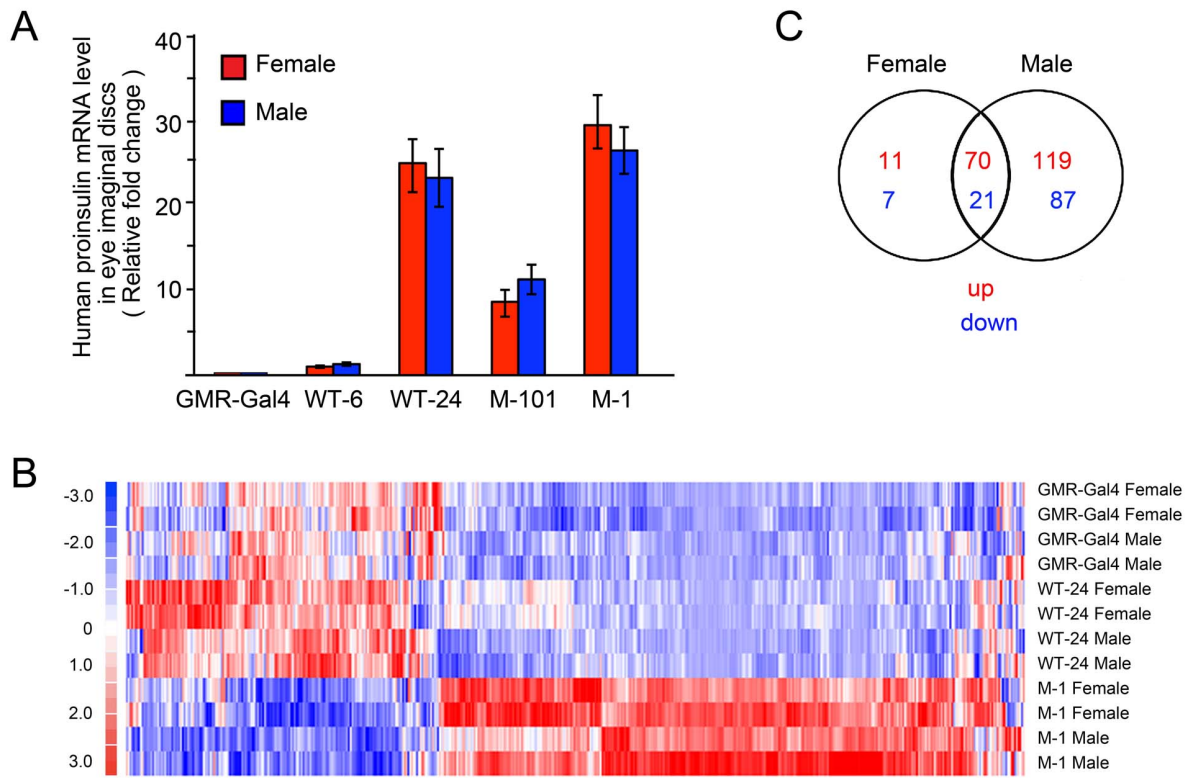

**Figure 2** Gene expression in eye-antennal imaginal discs of third-instar larva. (A) Relative mRNA levels in discs from larva expressing wild type (WT) and mutant (M, hINS$^{C96Y}$) human proinsulin. WT-6, WT-24, M-101 and M-1 are independent transgenic lines. Gene expression is normalized to the expression level of *rp49*. The values (mean ± SE) are shown relative to the ratio for female WT-6, set to one. (B) Heat map of expression profiles in rows (genes) and columns (lines x sex) for top 514 genes based on ANOVA between WT-24 and M-1 and the GMR-Gal4 control line are compared. ANOVA was performed for the three genotypes, two sexes, and two replicates according to the model $y = u + G + S + G \times S$, where $G$ is the genotype and $S$ is gender. Each gene was tested individually. The resulting P-values were ordered and false discovery rate (FDR, q) was estimated using Hochberg and Benjamini's method (HOCHBERG and BENJAMINI 1990). A list of 514 genes was selected to control FDR < 5%.

Each row is scaled to have mean 0 and variance 1. (C) Venn diagram showing the number of differentially expressed genes (up and down) in males and females in the comparison of WT-24 and M-1.

### Eye phenotype

The adult eye in GMR-Gal4 / UAS-hINS$^{C96Y}$ flies displays a number of characteristic defects, most notably a reduction in size, which varied among the transgenic stocks. Individual ommatidia are often collapsed, lacking the wild type organization of photoreceptor cells, giving the eye a glassy punctate phenotype (Figure 1 C, D). The regular array structure of ommatidia across the eye field is also disrupted, with the individual hairs projecting from each one either disarrayed or absent (Figure 1 G, H). Finally, black lesions can be present within the eye field where no cellular structure is evident (Figure 1 D, H). Mutant phenotypes require only a single copy of hINS$^{C96Y}$ expression (*i.e.*, GMR-Gal4 / UAS-hINS$^{C96Y}$ or UAS-hINS$^{C96Y}$, GMR-Gal4 / CyO) but are more severe in double dose (*i.e.,* UAS-hINS$^{C96Y}$, GMR-Gal4 / UAS-hINS$^{C96Y}$, GMR-Gal4).

The GMR-Gal4 driver activates expression in the cells posterior to the morphogenetic furrow in the eye discs (FREEMAN 1996). We therefore examined the organization and cellular structure of developing ommatidia by co-staining eye imaginal discs from late 3$^{rd}$ instar larvae with hINS anti-C peptide (a marker of proinsulin expression) (PARK *et al*. 2010) and an antibody against ELAV, a neuron-specific RNA-binding protein widely used to stain rhabdomeres (ROBINOW and WHITE 1991). This allowed us to confirm expression of wild type and mutant proinsulin in the developing eye field (Figure 1 K, L). Ommatidial arrays at this early stage of eye formation are irregular and disorganized (Figure 1 O, P), indicating that the adult reduced-eye phenotype originates in the eye morphogenetic furrow with improper formation and maturation of photoreceptor cells, ommatidia, and ultimately the entire eye field.

The severity of the reduced-eye phenotype differs quantitatively between the two sexes. Mutant males in GMR-Gal4 / UAS-hINS$^{C96Y}$ flies exhibit a measurably stronger *i.e.*, more degenerate, eye phenotype than females (Figure 1 C, G [female] vs. D, H [male]), a difference that is independent of gene dose, temperature, and genetic background (CARL 2010). This difference cannot be attributed to sex-specific difference in hINS expression, which does not differ significantly in the eye imaginal discs of either in wild type or mutant hINS lines (Figure 2A). The male biased phenotype, moreover, is not restricted to eye development; it is also observed in the notum and the wing when hINS[C96Y] is expressed in the developing wing imaginal disc under the control of three other Gal4 drivers, as described below. The sex-biased phenotype appears to arise, therefore, not through tissue-specific development, but rather through a gender difference in cellular response to the mutant insulin protein.

***Transcriptional profiles in eye imaginal discs expressing wild type and mutant hINS***

To reveal the effect of expressing hINS[C96Y] at the transcription level, and to identify the key changes in expression underlying the disease phenotype, we compared expression profiles of 3[rd] instar imaginal discs isolated from the GMR-Gal4 driver line and lines expressing wild type and mutant proinsulin, using the Affymetrix-GeneChip[®] Drosophila Genome 2.0 Array. The experiments included two independent transgenic lines each for hINS[WT] and hINS[C96Y], including the matched pair WT-24 and M-1 shown to express hINS mRNA at similarly high levels (Figure 2A). Compared to GMR-Gal4 control and accounting for sex differences, we found no evidence for any effect on gene expression by wild type hINS: no significant gene differences between the GMR-Gal4 control line and WT-6, and only a single difference in WT-24 at an FDR < 0.10 level (Figure S3). In contrast, 124 and 232 genes differed in males and females (respectively) between the mutant hINS lines and the GMR-Gal4 control. The effect on global gene expression caused by transgene expression can therefore be entirely attributed to the mutant proinsulin expression.

We then analyzed differences in gene expression between WT-24 and M-1, finding 297 genes whose expression differed in males (189 up-regulated and 108 down-regulated and 109 genes that differed in females (81 up-regulated and 28 down-regulated) (Figure 2C; File S1). Of these, 91 overlapped between males and females (70 up-regulated and 21 down-regulated). In contrast, using the same criteria we found only four genes showing significant differences between the control and hINS[WT] – expressing lines in both sexes. Another way to look at this is by fitting an ANOVA model to each gene for all three genotypes (M-1/WT-24/GMR-Gal4), accounting for sex effects (see Data analysis 2 in Materials and Methods), in which we identified 514 probe sets with significant genotype differences (File S2). A heat map (Figure 2B) illustrates the similarities between the transcription profiles of hINS$^{WT}$ and the control, and confirms at the molecular level the lack of a visible phenotype caused by hINS$^{WT}$ expression. It also highlights the reorganization of transcription induced by hINS$^{C96Y}$ expression.

Inspection of the genes whose expression changed in response to mutant hINS revealed genes involved in protein folding/modification, protein degradation, and defense response/programmed cell death (Table 2; Tables S2, S3), and include representatives in UPR and ER-associated degradation (ERAD) pathways. An unbiased and unsupervised clustering analysis using David tools for Gene Ontology (GO) terms showed the greatest enrichment in membrane-bound proteins, while heat-shock proteins were also enriched (Table S4).

Although we did not observe a significant difference in the mRNA levels of the upstream regulators of the UPR (Ire1, PEK(PERK), Hsc70-3 (BiP; GRP78) and Xbp1) in the GeneChip analysis, a more sensitive analysis by qRT-PCR in male eye imaginal discs expressing mutant hINS compared to GMR-Gal4 control revealed significant increases in expression of PERK (CG2087), BiP (CG4147), XBP1 (CG9415), and a marginally significant increase of IRE1 (CG4583; P = 0.08) (Table S5). As a more definitive test for activation of UPR, we also examined XBP1 mRNA for UPR-associated splicing by IRE1, and found evidence for it in mutant hINS-expressing cells but not wild type or GMR-Gal4-expressing cells (Figure S4). To confirm the microarray data by an independent method, we validated expression levels in the five lines (GMR-Gal4, WT-6, WT-24, M-101, M-1) for five genes sets (CG3966, CG7130, CG10420, CG10160 and CG9150) whose expression were up-regulated in males (Figure 2C). The results showed excellent correspondence between microarray and qRT-PCR (Table S6).

**Table 2 Selected genes up-regulated by GMR-Gal4 / USA-hINS[C96Y] in male eye imaginal discs**

| Probe set | Transcript | Name | Description (GO)[1] | Homolog[2] |
|---|---|---|---|---|
| **Protein modification/folding** | | | | |
| 1639033_at[3] | CG9432-RB | l(2)01289 | Disulfide isomerase | |
| 1623862_at[3] | CG3966-RA | ninaA | HSP[4] | |
| 1628660_at[3] | CG7130-RA | CG7130 | HSP[4] binding | DNAJB1 |
| 1623247_at[3] | CG10420-RA | CG10420 | HSP[4] | SIL1 |
| 1627525_a_at | CG1333-RA | Ero1L | Thiol-disulfide exchange | ERO1LB |
| 1641511_at | CG7394-RA | TIM14 | HSP[4] binding | DNAJC19 |
| 1641563_at | CG8286-RA | P58IPK | HSP[4] binding | DNAJC3 |
| 1634528_at | CG8412-RA | CG8412 | Glycosyltransferase | ALG12 |
| 1638456_at | CG8531-RA | CG8531 | HSP[4] binding | |
| **Protein degradation** | | | | |
| 1632071_at | CG8870-RA | CG8870 | Serine-type endopeptidase activity | |
| 1637515_s_at[3] | CG1512-RA | Cullin-2 | Ubiquitin-protein ligase | CUL2 |
| 1626272_s_at[3] | CG3066-RA | Sp 7 | Peptidase | SP7 |
| 1626460_at | CG2658-RA | CG2658 | Peptidase | SPG7 |
| 1635051_a_at | CG14536-RA | Herp | Ubiquitin-protein ligase | HERPUD2 |
| 1634486_at | CG30047-RA | CG30047 | Peptidase | |
| 1624372_at | CG10908-RA | Derlin-1 | Peptidase | DERL1 |
| 1637955_a_at[3] | CG1827-RA | CG1827 | Lysosome | |
| 1625253_at | CG4909-RA | POSH | Ubiquitin-protein ligase | SH3RF1 |
| 1623029_at | CG31535-RA | CG31535 | Ubiquitin-protein ligase | |
| 1634899_a_at | CG6512-RA | CG6512 | Peptidase | AFG3L2 |
| **Defense response/programmed cell death** | | | | |
| 1636668_at | CG9972-RA | CG9972 | Apoptosis[5] | |
| 1624450_at | CG6331-RA | Orct | Apoptotic process | |
| 1633145_at | CG4437-RA | PGRP-LF | Apoptosis[5] | PGLYRP3 |
| 1622979_a_at | CG7188-RB | Bax inhibitor-1 | Apoptosis[5] | |
| 1641298_at | CG10535-RA | Elp1 | Defense response | |
| 1634714_at | CG1676-RA | Cactin | Defense response | C19orf29 |
| 1635028_s_at | CG33047-RA | Fuca | Defense response | FUCA2 |
| 1638100_s_at | CG1228-RD | Ptpmeg | Apoptotic process | PTPN4 |
| 1628174_at | CG33119-RA | nim B1 | Defense response | |

[1] GO molecular function/process from http://www.flybase.org, www.uniprot.org and http://david.abcc.ncifcrf.gov
[2] Human homolog from http://flight.icr.ac.uk

### Expression of wild type and mutant hINS in the notum and wing

Expression of mutant (but not wild type) hINS in the notum, driven with an apterous driver (ap-Gal4), causes a reduction in the posterior margin of the notum and a loss of macrochaetae (Figure 3 C, D). The adult fly notum has 22 macrochaetae on the notum (Figure S2), which in ap>>hINS$^{C96Y}$ flies is reduced by an average of 8.3 and 13.4 bristles in females and males, respectively (Table S7). This 40% sex differential in bristle loss does not appear to be intrinsic to development — as a control we found no sex difference in bristle loss in the classic developmental mutant Scutoid (Sco) ([Fuse *et al.* 1999](#)), which suppresses notum bristles to approximately the same extent as mutant hINS expression, but does so to equal extent in both sexes (Table S7).

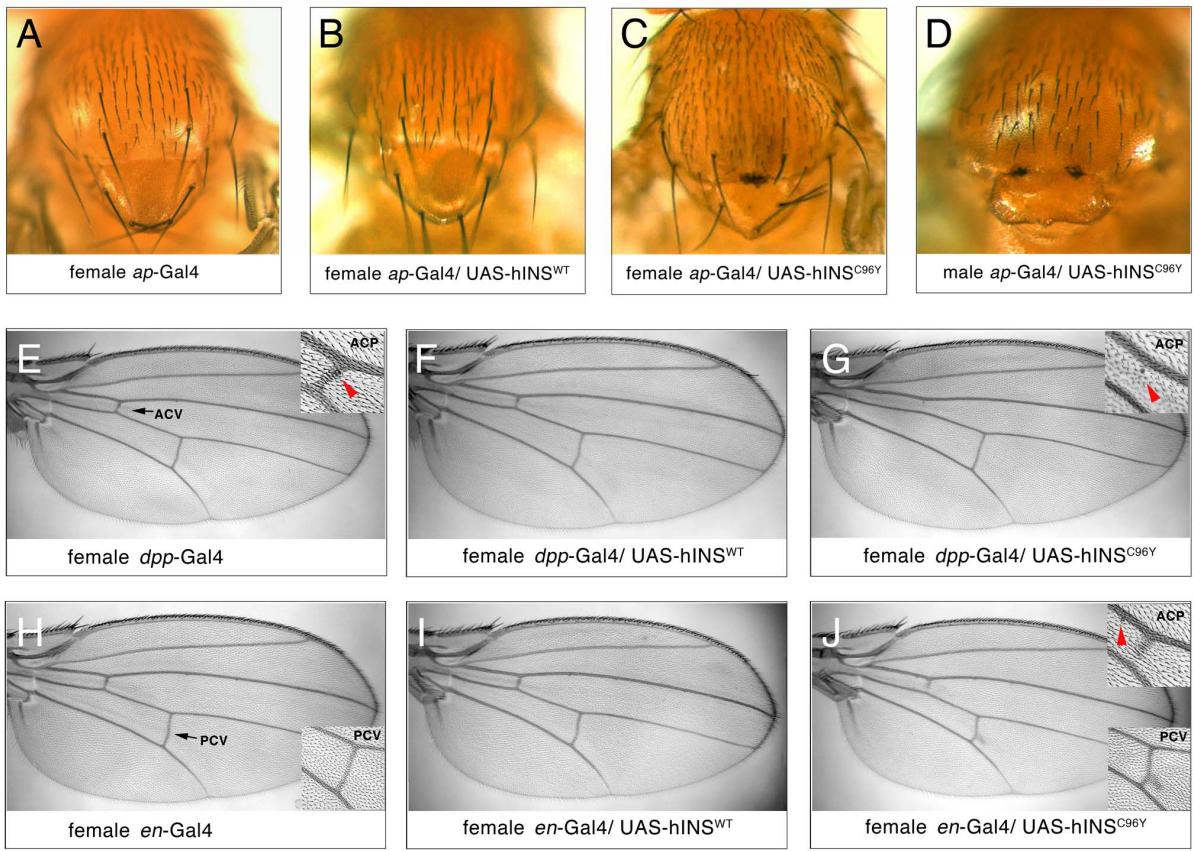

**Figure 3** Notum and wing phenotypes induced by hINS$^{C96Y}$ transgene expression. Notum (A-D) and wing (E-J) phenotypes in 3-5-day old adults of indicated sex and genotype. Insets show a higher magnification view of the anterior or posterior cross vein (ACV) with the *campaniform sensillae* shown by an arrow. Note the missing anterior cross vein in panel G (dpp-Gal4 driver), partial anterior cross vein and abnormal posterior cross vein in panel J (en-Gal4 driver), the relocation of the *campaniform sensillae* from the anterior cross vein to the longitudinal vein in panel J.

Expression of mutant (but not wild type) hINS in the developing wing imaginal disc causes visible defects in a proportion of adult wings (Figure 3 E-O). dpp-Gal4 drives expression in cells adjacent to the border of the posterior and anterior wing compartments; en-Gal4 drives expression only in the posterior wing compartment. In dpp-Gal4 / UAS-hINS$^{C96Y}$ flies, the distal margin of approximately 30% of wings are either scalloped or the anterior crossvein (ACV) absent, both phenotypes being restricted to the domain where mutant insulin is predicted to be expressed. Expression of mutant hINS by the en-Gal4 driver results in occasional partial loss of ACV along its posterior boundary, also corresponding to the predicted region of mutant protein expression.

Wing scalloping and ACV loss are striking phenocopies of the classical mutants *Notch* (incision of wing margin) and *crossveinless*, respectively, both regulators of wing development. Portions of the adult wing corresponding to mutant hINS expression in the wing imaginal disc are also significantly reduced in area (Figure 4).

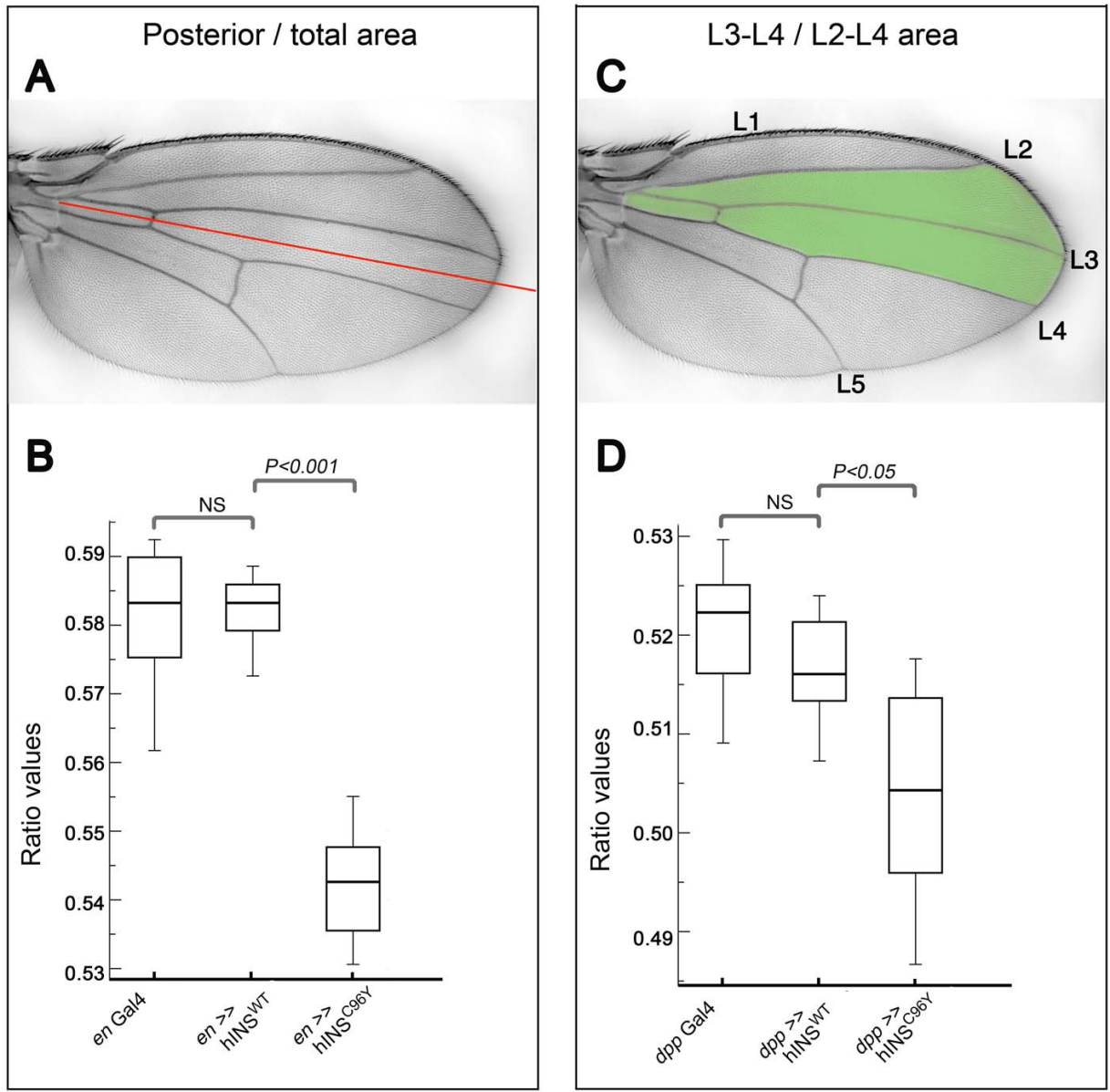

**Figure 4** Effect of hINS[C96Y] expression on female wing development. (A) Regions of en (red) and dpp (blue) expression. (**C**) en (genotype): en-Gal4 (n=13); en-Ga4 / UAS-hINS[WT] (n=13); and en-Gal4 / UAS-hINS[C96Y] (n=13). The values represent the ratio of the posterior wing compartment divided by the total wing area. (C) dpp (genotype): dpp-Gal4 (n=10), dpp-Gal4 / UAS-hINS[WT] ( n=10), and dpp-Gal4 / UAS-hINS[C96Y] (n=11) . The values represent the ratio of the L3-L4 intervein sector divided by the L2-L4 intervein sector area. NS, not significant Mann-Whitney U Test.

Mechano-sensory structures on the wing — the campaniform sensillae — can also be absent in portions of the wing where mutant hINS is expressed under the control of the dpp- and en-drivers (Figure S5). One such sensilla lies along anterior portion of ACV, and is typically absent when that portion of the crossvein is missing in en-Gal4 / UAS-hINS$^{C96Y}$ flies. In dpp-Gal4 / UAS-hINS$^{C96Y}$ flies, three additional sensillae sitting along the distal portion of the longitudinal wing vein 3, can also be absent (Table S8).

Mutant proinsulin expression in the developing wing also causes misspecification of cell fates to produce both ectopic wing veins and campaniform sensillae. The en-Gal4 driver, in particular, produces the novel appearance of both veins and sensillae (Figure 3 M, O). A sensilla sitting along the ACV, when absent in en-Gal4 / UAS-hINS$^{C96Y}$ wings, is often replaced with an ectopic one appearing more anteriorly along the ACV, or along the radial wing vein proximal to where it is intersected by the ACV. The posterior wing crossvein in the mutant can also project ectopic longitudinal veins.

### hINS$^{C96Y}$ induced phenotypes are modified by genetic background

The eye, wing and notum are notable examples of developmentally canalized structures that generally become more variable in a mutant background. Consistent with this observation, the mutant hINS-induced eye phenotype displays sensitivity to temperature and differs between the two sexes. In crosses involving the 3$^{rd}$ chromosome balancer TM3, we also observed more severe eye phenotypes when the balancer chromosome was present (CARL 2010), indicating sensitivity to the genetic background. We therefore examined the extent to which naturally occurring genetic variation modifies the mutant hINS phenotype in the eye and notum. We crossed a "tester" stock carrying the mutant transgene (M-1) and either the GMR-Gal4 or ap-Gal4 driver on the same 2$^{nd}$ chromosome (GMR>>hINS$^{C96Y}$ or ap>>hINS$^{C96Y}$) with 38 reference inbred lines derived from a single population collection, the DGRP (MACKAY *et al.* 2012) and measured eye phenotypes or counted dorsal macrochaetae in F$_1$ adults. The F$_1$ males in the crosses carried an identical X-chromosome — the tester chromosome. The screen, therefore, revealed only partially- or fully-dominant autosomal modifiers of the mutant phenotype. For each cross we measured eye area or dorsal bristle number in a minimum of 10 individuals of each sex.

*Eye phenotypes*: The crosses revealed highly heritable genetic variation ($h^2$ (males) = 0.732; $h^2$ (females) = 0.657), visible as a nearly continuous distribution of between-line differences in eye degeneration phenotypes ranging from nearly wild type to highly reduced and slit-like eyes (Figure 5 A, C). These inter-line differences are not correlated with either each line's body size (bi-variate fit of eye area with thorax length $r^2$= 0.0051 (Carl 2010)) or eye area (Figure S6), nor with the quantity of Gal4 protein, which did not vary significantly (Figure S7). There are also significant between-line differences in other aspects of the eye phenotype, including aspect ratio (width:height), ommatidial degeneration, and prevalence of lesions (Carl 2010). Lesion prevalence, unlike aspect ratio or ommatidial degeneration, was not significantly correlated with the extent of eye loss ($r^2$ = 0.01), indicating the two have independent genetic underpinnings rather than being the consequence of pleiotropy.

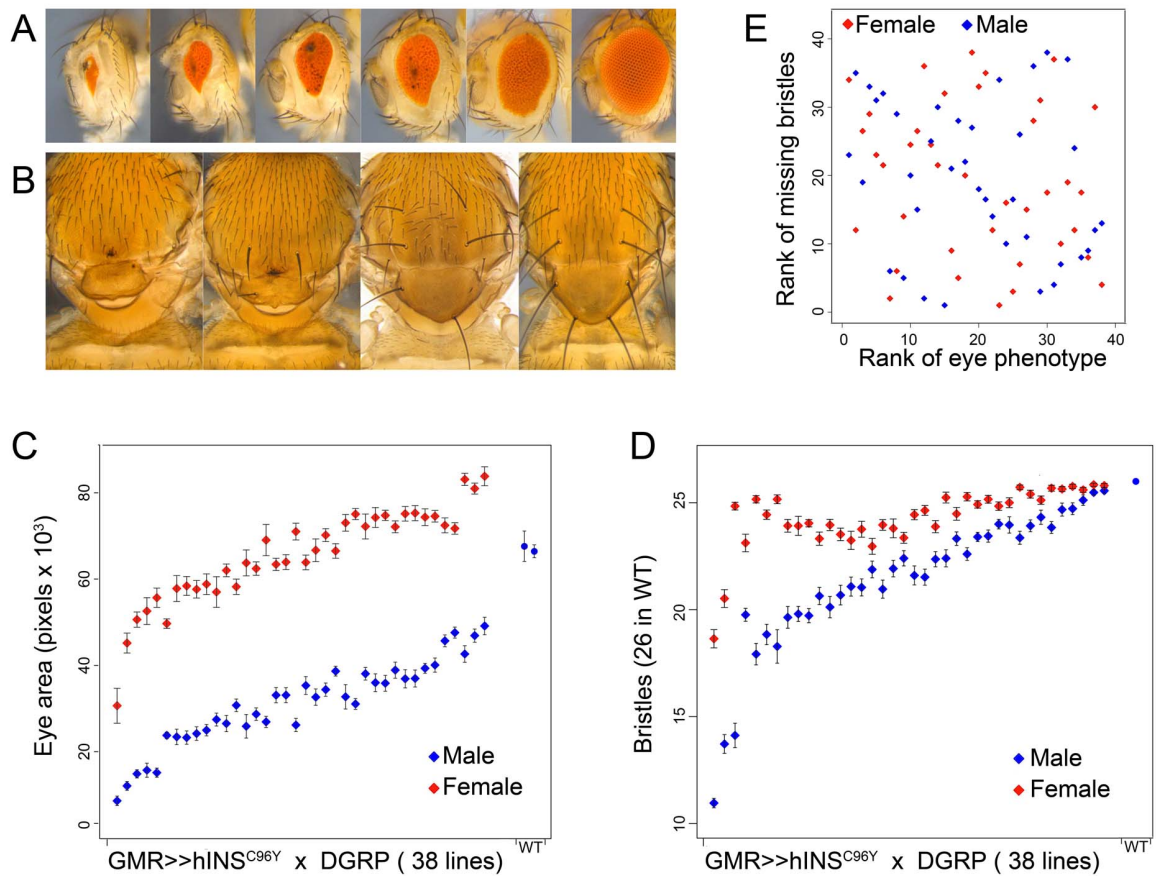

**Figure 5** Genetic variation for hINS$^{C96Y}$-induced degeneration in the adult eye and notum. (A) Variation in eye area in F$_1$ adults from crosses between GMR>>hINS$^{C96Y}$ tester strain and 38 DGRP lines described in Materials and Methods. (B) Bristle number in F1 adults from crosses in (A). (C and D) Eye area and bristle number. The data are displayed from left to right by decreasing severity of phenotypes. Eye area (mean ± SE)

for a wild type control (GMR-Gal4) is shown on the far right in filled circles (in C, only male wild type eye areas are shown). (E) Correlation between bristle loss and eye-area reduction (male: Spearman's rank correlation $rho$ = -0.23, P = 0.16; female: $rho$ = -0.17, P = 0.30).

*Notum phenotypes*: As with the eye phenotype, we found significant inter-line variation ranging from lines with nearly wild type bristle number (RAL-427, bristle reduction = 0.44 ± 0.65) to ones missing a majority of bristles (RAL-335, bristle reduction = 15.04 ± 1.10) and a high heritability ($h^2$ (males)=0.744).

### Disease traits are uncorrelated

We reasoned that if the genetic pathways responding to hINS[C96Y] expression common to both eye and notum, such as UPR, harbor modifiers of the response, then the severity of the eye reduction and bristle loss should be positively correlated in the sample of DGRP lines. To ask whether the same modifiers are acting in similar manner in both tissues we measured the correlation between traits in the 38 lines for which both were measured. Surprisingly, we found no evidence for a positive correlation (Figure 5E; Table S9; Male: Spearman's rank correlation $rho$ = -0.23, P = 0.16; Female: $rho$ = -0.17, P = 0.30). The common response pathways either harbor little of the genetic variation for the disease phenotypes, or their penetrance must be modulated by tissue-specific factors.

### Variation in eye-specific genetic pathways is uncorrelated with hINS[C96Y]–induced phenotypes

The lack of correlated mutant hINS induced phenotypes in the eye and notum raises an alternative possibility that genetic variation acts not through shared response pathways but rather through tissue-specific developmental pathways, and in so doing "releases" pathway-specific genetic variation otherwise suppressed in the wild type. To test this possibility, we examined genetic variation in the DGRP lines for two eye-development-specific genetic mutations *Lobe* (*L*) and *Drop* (*Dr*). *L* and *Dr* are classic dominant eye-degeneration mutations that can be crossed to the DGRP lines in the same manner as the mutant insulin transgene to reveal dominant genetic variation for reduced-eye phenotypes. *L* encodes the ortholog of mammalian PRAS40 and regulates eye development through TORC1 signaling (WANG AND HUANG 2009); mutants display an apoptotic reduced-eye phenotype. *L* acts through the Jak/Stat signaling pathway in the ventral eye, possibly interacting with the *Notch* ligand, *Serrate* (CHERN AND CHOI 2002). *Dr,* in contrast, is a muscle segment homeobox-1 transcription factor that regulates interaction between epithelial and mesenchymal cells. It is active in embryonic neural dorsal-ventral patterning and again in eye development. *Dr* mutants ectopically express the *muscle segment homeobox* (*msh*) gene, blocking morphogenetic furrow progression in the developing eye, leading to apoptotic photoreceptor cell loss and a nearly eyeless phenotype (MOZER 2001).

The genetic variation exposed by mutant hINS expression appears to be distinct from the genetic variation exposed by eye development mutants despite its apparent tissue specificity. We crossed *L, Dr* and hINS$^{C96Y}$ to 38 DGRP lines and collected F$_1$ adults for eye area measurement. Variation in hINS$^{C96Y}$-induced eye degeneration was comparable to previous measurements in the same lines (Figure 6A, D; Figures S1, S8; Table S10, S11). F$_1$ flies displayed heritable variation for both *Dr* and *L* phenotypes, which when scaled by their within-line variances displayed a range of phenotypes similar to hINS$^{C96Y}$ flies (Figure 6A, C). There is no significant correlation between any pair of traits (Figure 6C; Table S9), and thus no evidence for shared variation acting on mutant hINS and two eye development specific mutants. It is also worth noting that for both *L* and *Dr,* eye area in males is approximately 85% that of females, consistent with the difference in wild type flies. In contrast, eye area in males of mutant insulin-expressing crosses is 50% that of females, indicating a sex-specific input to the disease phenotype (also see Figures 5C and D).

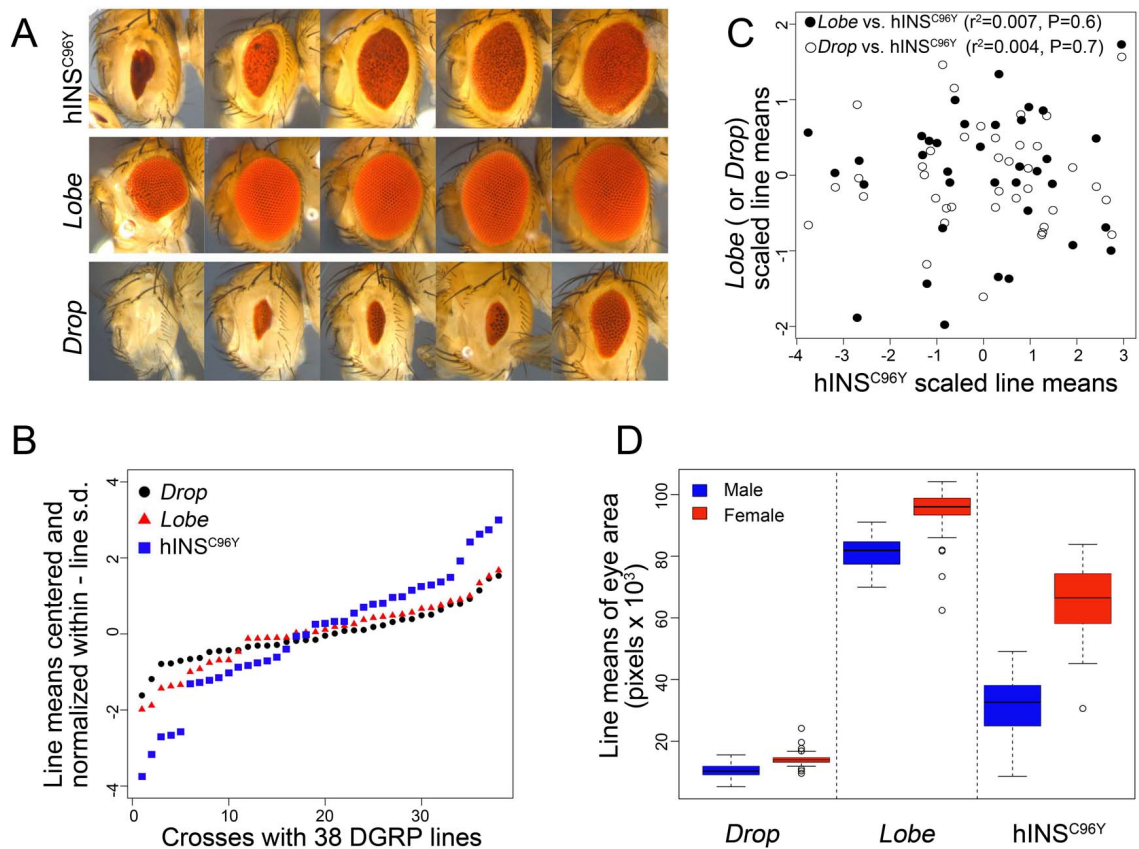

**Figure 6** Genetic variation for eye area reduction in $F_1$ adults from crosses between GMR>>hINS^C96Y, *Drop*, or *Lobe* and 38 DGRP lines. (A) Range of phenotypes in $F_1$ adults in both sexes. (B) Deviation (in units of within-line SD) of each line mean from the overall mean within each of the three sets of crosses. (C) Correlation of eye area reduction between hINS^C96Y and *Dr* (open circle) or *L* (closed circle) x DGRP $F_1$ males. (D) Boxplots showing the unscaled distribution of phenotypes in the three sets of crosses (thick line: median; box: 25th and 75th percentile; whisker: 1.5 interquartile range (IQR); circles: data outside the 1.5 IQR).

**DISCUSSION**

*Drosophila* is a useful model for studying cell function and development in response to misfolded proinsulin. We show that mutant (but not wild type) hINS expression causes a reduction in size (and cell number for eyes) in every tissue examined. Human proinsulin is not processed to insulin in developing eye cells, but can be induced to do so by overexpressing a secretory cell master regulator, the bHLH transcription factor DIMMED (PARK *et al.* 2012). Consistent with this result, we observed no effect of wild type hINS expression on gene expression or eye development. Although we have not established specific mechanism(s) by which mutant-induced eye reduction occurs, one of them is likely to involve UPR, which we show is induced based on both the presence of Xbp1 alternatively spliced mRNA in eye imaginal discs expressing hINS$^{C96Y}$, and on the induction of well-known stress response genes, including ones aiding protein folding and promoting programmed cell death. We also establish a reorganization of gene expression in imaginal disc cells in response to mutant hINS expression.

Cell death in the *Drosophila* model recapitulates a key feature of disease observed in mouse diabetes caused by the same C96Y mutation in *Ins2*: the dominant loss of insulin-secreting beta cells (KAYO and KOIZUMI 1998; WANG *et al.* 1999). In the mouse model, the synthesis of misfolded proinsulin leads to its retention in the ER resulting in induction of UPR, death of the insulin-secreting pancreatic beta cells, and diabetes (SONG *et al.* 2008; TABAS and RON 2011). The human form of hINS$^{C96Y}$-induced disease is believed to act through the same mechanism (LIU *et al.* 2010; PARK *et al.* 2010); based on our gene expression experiment, this may hold true in the *Drosophila* model as well.

Developing tissues, we discovered, are more sensitive to mutant hINS expression in males than females. When expressed in the eye, hINS$^{C96Y}$ causes a nearly two-fold reduction in eye area in males compared to females. Other features of the eye, including the presence of necrotic lesions, photoreceptor cell collapse, and ommatidial disorganization, are also more evident in males. *L* and *Dr* in contrast, although also producing reduced-eye phenotypes, do not exhibit sex-specific differences relative to wild type. The flexibility of the *Drosophila* model allowed us to establish that the notum also displays a differential male sensitivity to mutant hINS expression. We believe, therefore, that the greater sensitivity to mutant hINS in males must involve cell physiology rather than tissue-specific development. We can entertain at least two hypotheses for the male sensitivity, both of which are potentially testable. One obvious possibility involves disruption of dosage compensation. In *Drosophila*, dosage compensation occurs in males by up-regulating X-linked genes through the activity of the male sex lethal (MSL) complex (GELBART and KURODA 2009). Reorganization of gene expression in stressed cells may disrupt maintenance of dosage compensation, leading to the exacerbation of cellular stress and cell death in males. An alternative hypothesis posits that cells in males are less well canalized against perturbation, such as with expression of mutant hINS, perhaps because dosage compensation introduces greater variability in X-linked gene expression. It is well known, for example, that the effectiveness of dosage compensation varies quantitatively across X-linked genes and is complete in only a subset of them (HAMADA *et al.* 2005). Cell-to-cell or temporal variation in X-linked gene expression might increase demand on the homeostatic mechanisms involving proteostasis. It should be possible to test these hypotheses by genetically manipulating flies to examine sex determination, dosage compensation, or sex differentiation pathway contributions to male biased disease. More generally, fly models of human disease, such as ours, may be valuable in disentangling environmental and genetic contributions to sex differences in susceptibility or severity of disease, a notoriously difficult problem in human studies.

Male sex bias may be a general property of the disease: it is also a feature of diabetes in mice (WANG *et al.* 1999). Male mice heterozygous for *Ins2*$^{C96Y}$ develop diabetes at an earlier age than females (OYADOMARI *et al.* 2002). In the fly, X-linked genes are up-regulated in males whereas in mammals a single X-chromosome is inactivated in females cells. If the mechanism underlying the male bias in fly and mouse is the same, it is unlikely, therefore, to directly involve dosage compensation.

A second unexpected finding was the presence of fully differentiated ectopic veins and sensory structures in wings expressing mutant hINS. These same wings also display loss-of-structure phenotypes, including crossveins and campaniform sensillae, as well as scalloping of wing margins. Both ectopic gain and loss of these differentiated tissues are striking phenocopies of classical wing mutations, many of which have been shown to be involved in the regulation of wing development (Neto-Silva *et al.* 2009). We believe, therefore, that mutant hINS expression can not only induce cell death, but can also lead to reprogramming of cell fates. An interesting implication for the human form of the disease is that loss of beta cells in neonates may involve not only cell death but also transformation of precursor cells to other cell types.

Third, crosses to a reference panel of naturally derived lines (DGRP) revealed extensive dominant (or partially-dominant) genetic variation acting to suppress or enhance cell loss. One possibility, which we investigated and could reject, is variation in mutant hINS gene expression in different DGRP backgrounds. Since all the flies carry the same tester chromosome, (GMR>>hINS$^{C96Y}$), we focused our attention on Gal4 instead, because its expression could be influenced by variation in transcription factors acting on its promoter, GMR; we found no evidence for differences in Gal4 protein levels between DGRP lines representative of the full range of eye degeneration phenotypes (Figure S7). GMR is a synthetic enhancer consisting of binding sites for the eye-specific transcription factor Glass (*Gl*). In an accompanying paper, we also find no evidence for association of genetic variation in or around the *Gl* locus with eye degeneration (He *et al.* 2013). We do not believe, therefore, that variation in eye degeneration is caused by genetic variation in the transcription of mutant hINS.

Finding extensive genetic variation in eye degeneration in our $F_1$ screen establishes the feasibility of applying methods of statistical association to identify modifiers of disease, the subject of the accompanying paper (He *et al.* 2013). Here we explored other dimensions of this variability. It is worth noting that many, if not most, Mendelian models of disease in the fly involve gain-of-function alleles, which facilitates screens for natural variation in $F_1$ flies. In additional to the convenience of this genetic screen, it also eliminates phenotypes resulting from the homozygosity of deleterious alleles in inbred lines. Outcrossed genotypes are well suited for investigating low frequency variants, which are rarely homozygous in natural populations.

Disease phenotypes in the eye and notum were not significantly correlated in the DGRP panel, suggesting that different suites of alleles are acting in the two tissues. A positive correlation would be expected if genetic variation occurred primarily in shared pathways responding to mutant hINS expression, such as UPR. Not finding evidence for such a correlation, we then investigated whether a correlation would be observed comparing a single phenotype — eye reduction — caused by hINS and by two classical mutations. The fact that we failed to find significant correlations between either *L* or *Dr* and hINS$^{C96Y}$ leaves us with a puzzling set of results: natural variation for hINS-induced disease severity exhibits tissue specificity but involves a different set of genes or alleles than ones revealed with eye-development-specific mutants. The latter result, but perhaps not the former, should come as no surprise. In other models of Mendelian disease, e.g. aggregation-prone proteins expressed in the developing eye, forward genetic screens for suppressors and enhancers of reduced eye phenotypes successfully identify genes acting in pathways known to be responsive to proteostatic stress, UPR, apoptosis, RNA-folding, and peptide folding, transit, and degradation pathways (Chai *et al.* 1999; Warrick *et al.* 1999; Chan *et al.* 2000; Chan and Bonini 2003; Bilen and Bonini 2005; Warrick *et al.* 2005; Bilen and Bonini 2007; Lessing and Bonini 2008; Li *et al.* 2008; Yu and Bonini 2011), but not regulators of eye development. As this also appears to be the case for naturally occurring variation in our Mendelian model of disease, distinct alleles and genes must be acting as modifiers, perhaps epistatically, in different tissues.

An alternative hypothesis can be constructed on the premise that the spectrum of mutations affecting this complex disease trait may have a much broader set of targets, needing only to impinge on processes involved in cellular or physiological homeostasis. Disease occurs when an individual's homeostatic "capacitance" – the ability to buffer against cellular stress – is exceeded. Whether a threshold is crossed will depend on both the cellular activities set by an individual's background genotype and on environmental demands or rare mutant alleles acting in critical pathways.

Subtle effects of genetic background on the ability of a cell to balance protein synthesis, folding, transport, and degradation —*i.e.*, proteostasis — may be responsible for many diseases, in addition to diabetes. Under this hypothesis, a complex and diffuse web of interacting polymorphisms set an individual's ability to respond to genetic or environmental challenges, determining susceptibility to and severity of disease. If true, the vast majority of mutations and the spectrum of disease-causing loci segregating in natural populations are likely to be systematically and substantially different than the strong loss- or gain-of-function alleles identified in forward genetic screens alone.

In addition, as proteomes differ between tissues, so too will the alleles affecting proteostasis. This possibility is illustrated by revealing experiments on two aggregation-prone / misfolded proteins in a worm model: polyglutamine protein ((GIDALEVITZ *et al.* 2006) and mutant SOD1 (GIDALEVITZ *et al.* 2009). In both cases, temperature sensitive (*ts*) mutations in housekeeping proteins, although innocuous when the worm is reared below the *ts* threshold, enhance mutant protein phenotypes, and hence toxicity, when the *ts* threshold is exceeded. SOD1 phenotypes are also sensitive to the genetic background.

*Drosophila* is an excellent model for investigating naturally occurring genetic variation for quantitative traits. The recent establishment of the DGRP (MACKAY *et al.* 2012), synthetic populations (HUANG *et al.* 2012; KING *et al.* 2012), and other novel population resequencing approaches (TURNER and MILLER 2012), add to its power and appeal. Here we extend the applicability of these approaches to the study of human disease. An important question remaining to be addressed is whether the extensive genetic variation revealed in this study of a genetically "sensitized" fly is the same as the variation underlying complex genetic forms of the disease, an issue further discussed in the accompanying paper. Affirmative answer to this question raises the prospect for using *Drosophila* as a model of genetically complex human disease.

## ACKNOWLEDGMENTS

This work was funded by grants from the National Institute of Diabetes and Digestive and Kidney Diseases (R01DK013914 and P30DK020595), the National Institute of General Medical Sciences (P50GM081892), the Chicago Biomedical Consortium with support from the Searle Funds at The Chicago Community Trust, and a gift from the Kovler Family Foundation. We thank Honggang Ye, Esme Gaisford, Amanda Neisch, Richard Morimoto, Ilya Ruvinsky, Richard Hudson, Rick Fehon and Ilaria Rebay for technical help and advice. The content is solely the responsibility of the authors and does not necessarily represent the official views of the National Institute of Diabetes and Digestive and Kidney Diseases, the National Institute of General Medical Sciences or the National Institutes of Health.

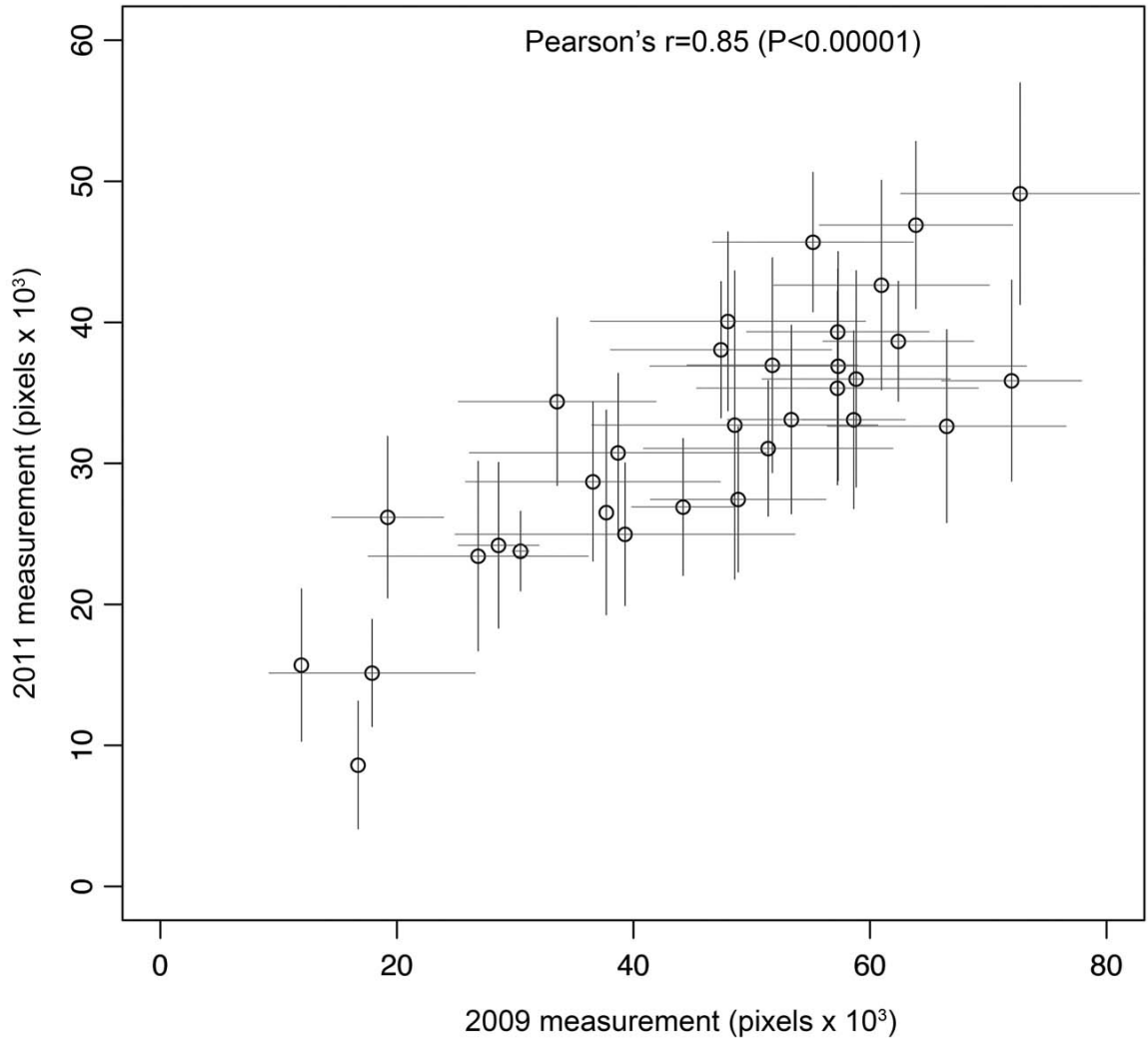

**Figure S1** Correlation of GMR>>hINS$^{C96Y}$ eye phenotypes in crosses to DGRP lines. Shown are data collected in two independent experiments carried out in 2009 and 2011. Plotted are average eye areas for approximately 10 males, as described in the Materials & Methods.

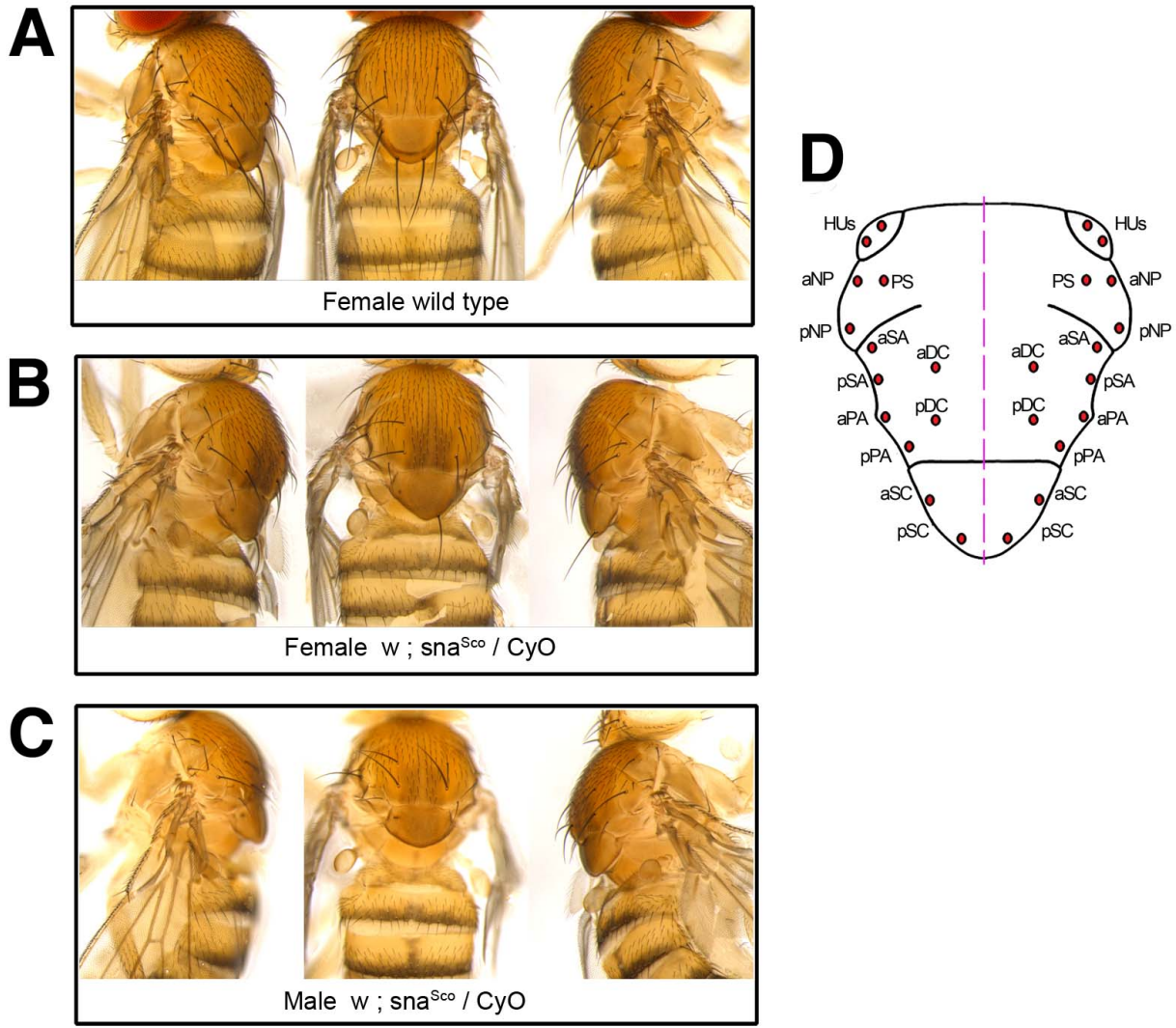

**Figure S2** Bristle count. Notum bristles (macrochaete). (A) Wild type female. (B), (C) Notum of Sco mutant (w1118; CyO dfd-YFP / snaSco); (B) female; (C) male. (D) Macrochaete positions on heminotum and humerus with their nomenclature. aDC, anterior dorsocentral bristle; aNP, anterior notopleural bristle; aPA, anterior postalar bristle; aSA, anterior supraalar bristle; aSC, anterior scutellar bristle; HU, humeral bristle. Humeral bristles are prothoracic structures that differentiate from first leg imaginal discs; they were always present in the ap>>hINS$^{C96Y}$ crosses, which is expressed in the mesothorax. Sco mutants, in contrast, often lack humeral bristles, as well as bristles typically lost in ap>>hINS$^{C96Y}$ flies.

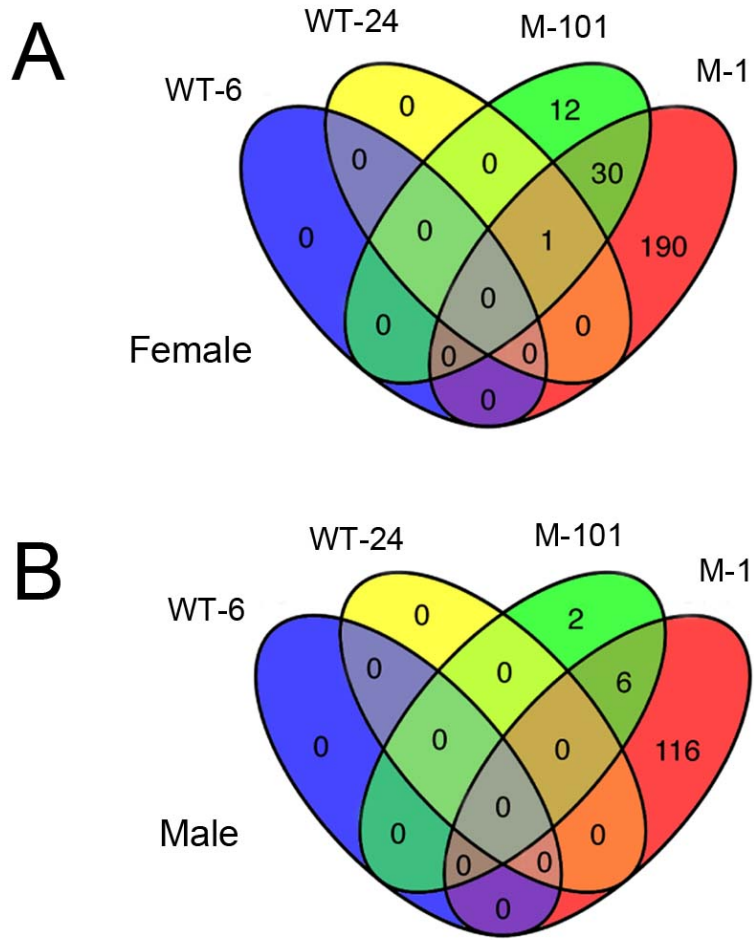

**Figure S3** Comparison of gene expression in 3$^{rd}$ instar eye imaginal discs from GMR-Gal4 and GMR-Gal4 / UAS-hINS$^{WT}$ (Lines WT-6 and WT-24) and GMR-Gal4 / UAS-hINS$^{C96Y}$ (Lines M-101 and M-1). (A) Venn diagram showing differential expression between GMR-Gal4 background and hINS transgenic lines in female larva. (B) Venn diagram showing differential expression between GMR-Gal4 background and hINS transgenic lines in male larva. Expression data were analyzed by sex with one-way ANOVA; significant genes were then tested to determine whether mean expression of hINS$^{WT}$ or hINS$^{C96Y}$ was significantly different from the GMR-Gal4 control (see Data analysis 1 in Material and Methods).

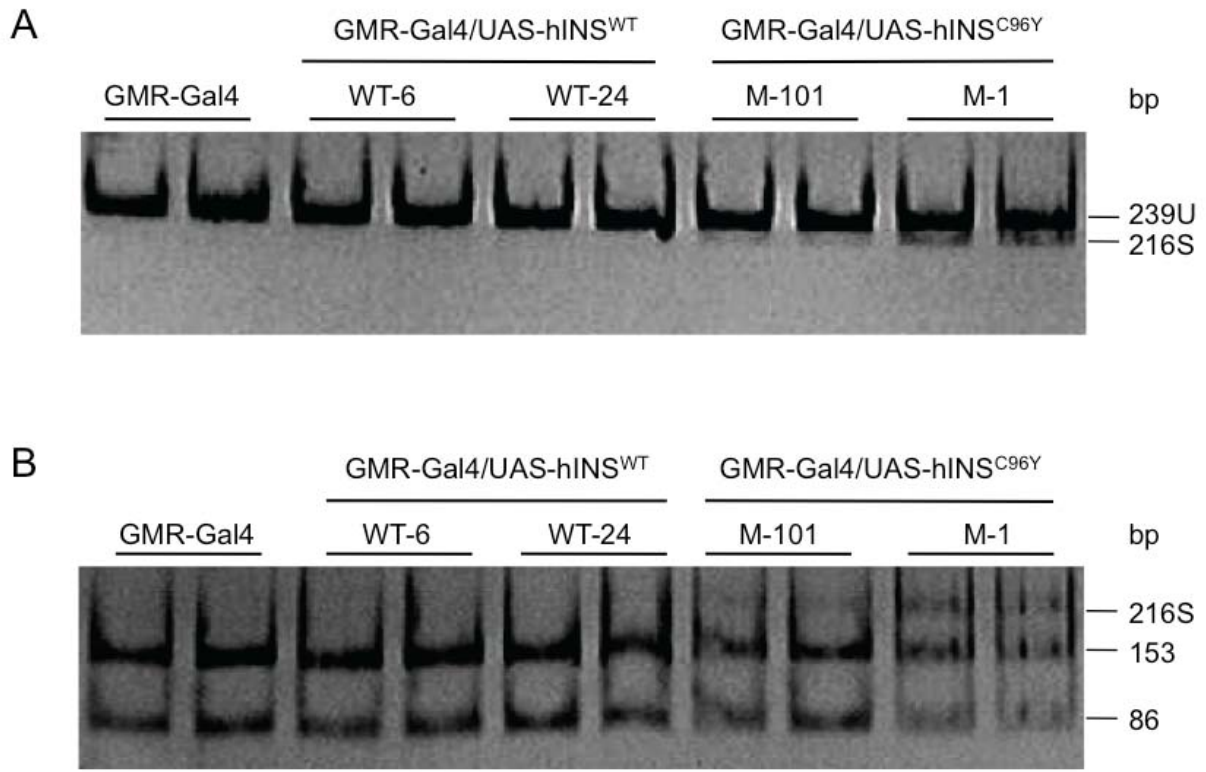

**Figure S4** Alternative splicing of Xbp1 in RNA. RNA was isolated from eye imaginal discs of 3$^{rd}$ instar larva of indicated genotype. (A) RT-PCR and 10% PAGE analysis of the expression of the Xbp1-unspliced (U) and Xbp1- spliced (S) transcripts. Unspliced and spliced isoforms could be distinguished by the size of the PCR product (239 and 216 bp, respectively). The PCR products were detected by ethidium bromide staining. (B) To further resolve the two isoforms, the PCR products were treated with *Pst*I to cleave the unspliced form into fragments of 153 and 86 bp.

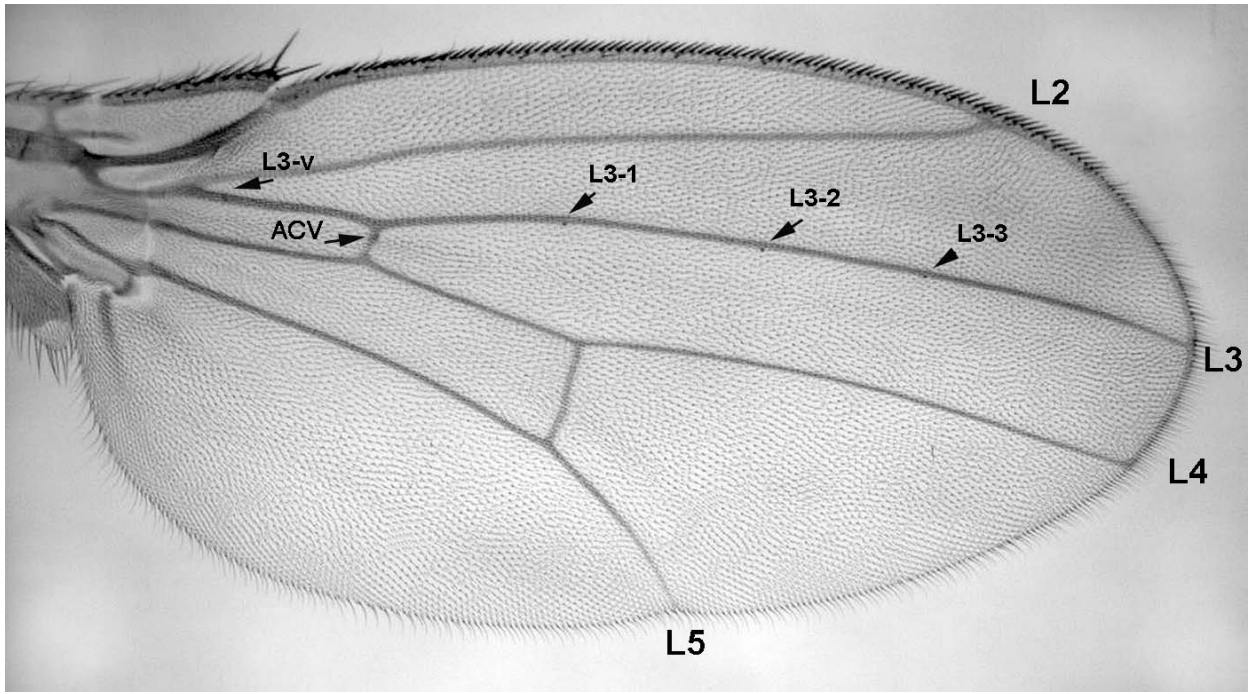

**Figure S5** Campaniform sensilla on the wing: L3-v, ACV, L3-1, L3-2, L3-3 (black arrows). L2, L3, L4, and L5:  longitudinal veins of the wingblade are numbered.

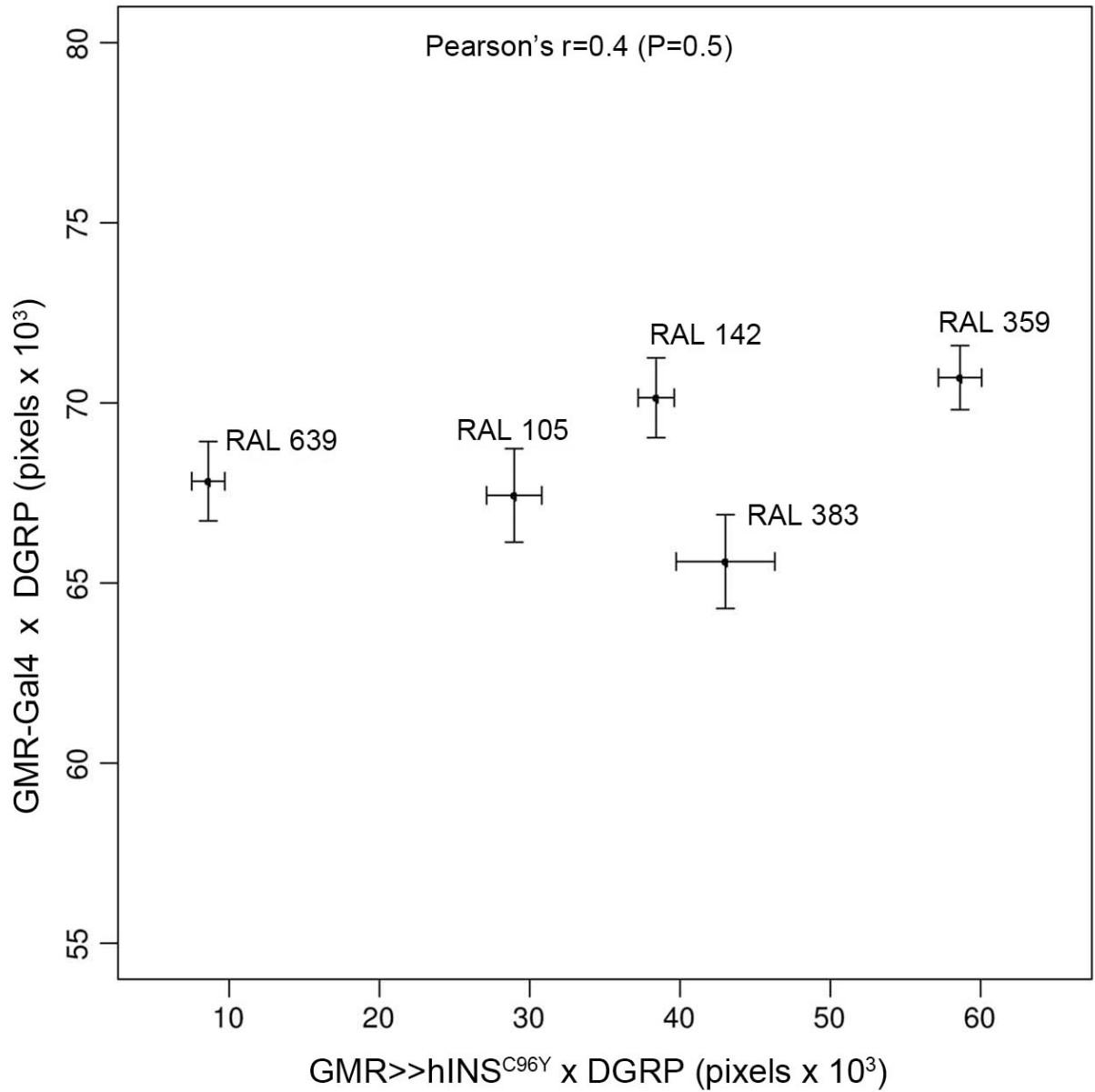

**Figure S6** Eye area in crosses of five DGRP lines to hINS[C96Y] is not correlated with wild type eye area. Five DGRP lines were sampled across the phenotypic distribution of the crosses with hINS[C96Y], including the two extremes. They were crossed to a control line (GMR-Gal4), whose male progeny were measured for their eye area. No correlation is observed between results from the hINS[C96Y] cross and the GMR-Gal4 cross.

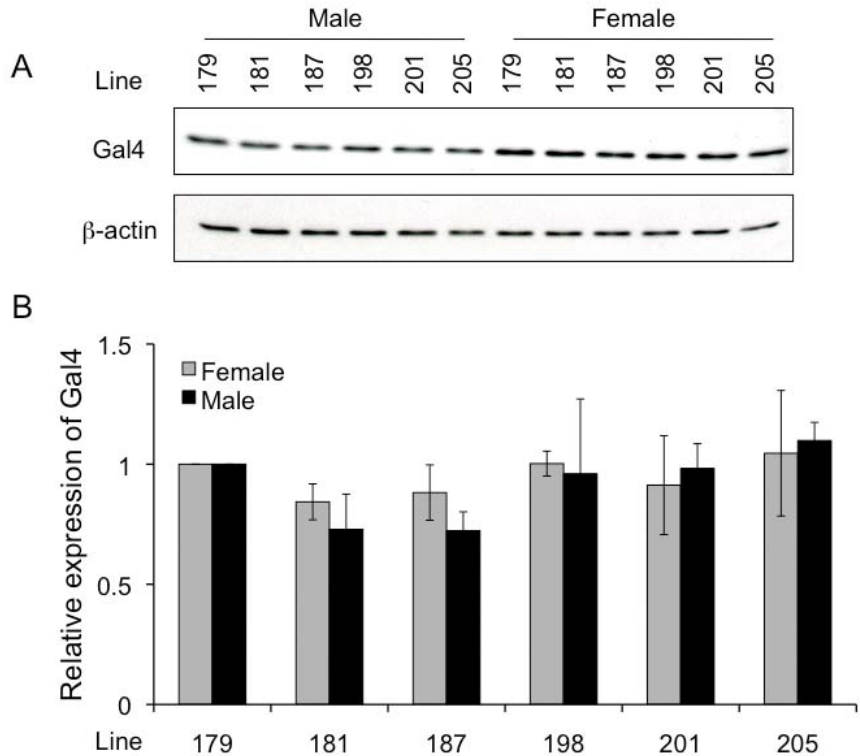

**Figure S7** GAL4 concentrations do not differ between DGRP lines in crosses to GMR>>hINS[C96Y]. (A) Western blot showing Gal4 bands in crosses between GMR>>hINS[C96Y] and five DGRP lines that were selected to span the range of eye phenotypes, as shown in Figure 5. (B) Mean value of Gal4 expression in two technical replicates. The density of the Gal4 band was normalized to β-actin (control); values shown are the fold change relative to male and female line 179.

Method: Ten μg of total protein from cell lysates prepared from 20 adult heads was separated on a 10% SDS-PAGE, transferred to a PVDF membrane (Amersham Hybond™-P PVDF Transfer Membrane; GE Healthcare, Piscataway, NJ), and incubated with rabbit polyclonal anti-Gal4 primary antibody (Santa Cruz, CA; sc-577; 1/1,000 dilution and a donkey anti-rabbit IgG-HRP secondary antibody (Santa Cruz, sc-2096; 1/5,000 dilution). The blot was developed with Amersham ECL™ Western Blotting Detection Reagents and detected by chemiluminescence. The blot was also probed with a mouse β-actin antibody (Santa Cruz; sc-47778; 1/1,000 dilution) as a loading control. Band intensity was quantified using the Gel Analysis package in ImageJ Software (NIH).

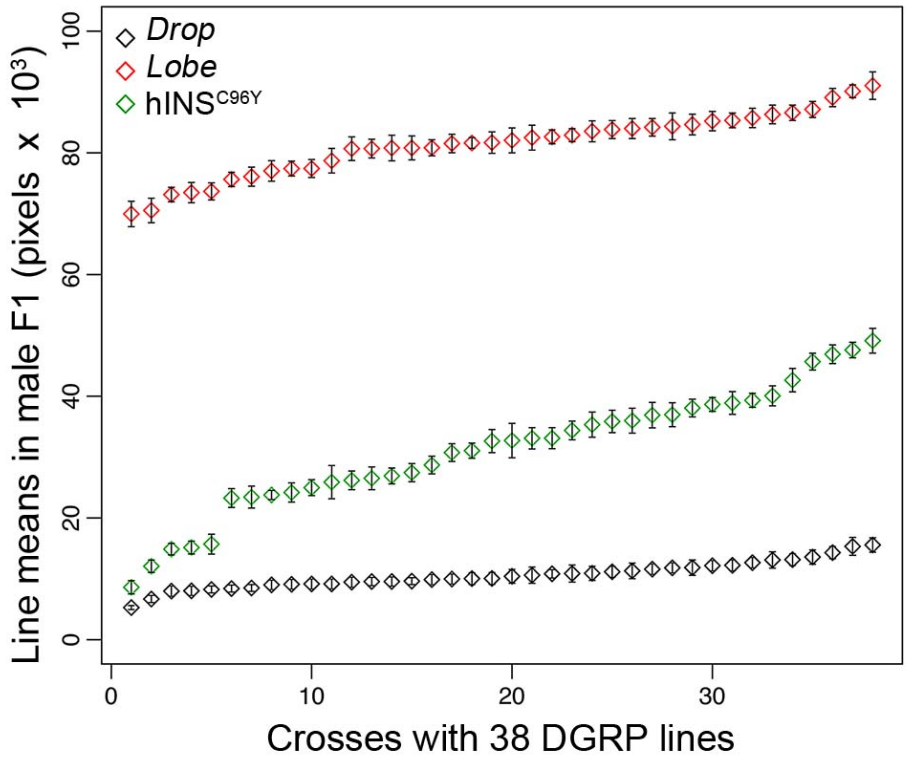

**Figure S8** Untransformed eye areas in $F_1$ adults produced from crosses of *Dr*, *L*, or GMR>>hINS[C96Y] to DGRP lines. Data shown are mean ± SE. Although *Dr* and *L* eye area varies less than GMR>>hINS[C96Y], the between-line (i.e., heritable) differences are comparable when scaled by within-line variances (Figure 6).

**Table S1 Primer sequences used for quantitative RT-PCR**

| Gene | Forward | Reverse |
|---|---|---|
| Human proinsulin | CTACCTAGTGTGCGGGGAAC | GCTGGTAGAGGGAGCAGATG |
| CG4583 (Ire1) | GAGATCACAGCGAACGACAA | GGATAATTCGGCTGTCCTCA |
| CG2087 (PEK) | GTGGTTCTGGTGGAAGGAAA | GGCACATGACGTTCAATGAC |
| CG4147 (Hsc70-3) | CAAGTTCGAGGAGCTCAACC | AATCTCGTGCACGTCCTTCT |
| CG9415 (Xbp1) | AGAACCACAAGCTGGACTCG | CAGATCCAAGGTTGGTGGAC |
| CG3966 | TGGCTGCCAGTTTTATGTGA | CGGGTAGAACTCGAACTGCT |
| CG7130 | ACAAGATTCTGGGCATCGAG | CGCGCTTTTCCTTATCAAAG |
| CG10420 | GGAGGCAAGACAAGCTGAAG | TAGCTTGACCTTCCGCAATC |
| CG10160 | TTTAGAGGCGCCCAAAATAA | GAGACGTTCTGAGCCAGGAT |
| CG9150 | CGAAGGTCACGTTCTCATCA | TAACCCGGATTTTGTTCGAG |

**Table S2  Selected genes up-regulated by GMR-Gal4 / UAS-hINS[C96Y] in male eye imaginal discs**

| Probe set | Transcript | Name | Description (GO)[1] | Homolog[2] |
|---|---|---|---|---|
| **Transport** | | | | |
| 1634512_at[3] | CG5226-RA | CG5226 | Carnitine transport | SLC6A17 |
| 1637439_at | CG14709-RA | Mrp4 | Transport | ABCC4 |
| 1635700_at[3] | CG31792-RA | CG31792-RA | Ion transport | |
| 1627582_a_at | CG30035-RA | Tret1 | Trehalose transport | SLC2A8 |
| 1624450_at | CG6331-RA | Orct | Ion transport | |
| 1623247_at[3] | CG10420-RA | CG10420 | Intracellular protein membrane transport | SIL1 |
| 1636800_at[3] | CG13610-RA | Orct2 | Ion transport | |
| 1641606_s_at[3] | CG6608-RB | Tpc1 | Thiamine pyrophosphate transport | |
| 1632622_ | CG32538-RB | GfA | Ion transport | |
| 1629040_at | CG3476-RA | CG3476 | Ion transport | |
| 1633039_at[3] | CG5646-RA | CG5646 | Mitochondrial transport (acyl carnitine) | |
| 1627945_at[3] | CG4205-RA | Fdxh | Electron transport | FDX1L |
| 1625250_at | CG5802-RA | CG5802 | Sugar transport | SLC35B1 |
| 1630804_at* | CG6417-RA | Oatp33Eb | Ion transport | |
| 1633536_at | CG4630-RA | CG4630 | Transport | |
| 1635684_a_at | CG2999-RA | unc-13 | Synaptic vesicle exocytosis | UNC13C |
| 1641511_at | CG7394-RA | TIM14 | Membrane transport | DNAJC19 |
| 1623743_at | CG3191-RA | CG3191 | Transport | |
| 1631763_at | CG31793-RA | CG31793 | Transport | ABCC4 |
| 1637280_at | CG4861-RA | LpR1 | Receptor-mediated endocytosis | VLDLR |
| 1640075_a_at | CG3424-RA | pathetic | Transport | |
| 1640220_a_at | CG11779-RA | CG11779 | Transport | TIMM44 |
| 1633304_at | CG1967-RA | p24-1 | Post-Golgi vesicle-mediated transport | TMED7 |
| 1632676_s_at[3] | CG11897-RA | CG11897 | Transport | |
| 1637772_at | CG4726-RA | MFS3 | Ion transport | |
| 1631856_a_at | CG7361-RB | RFeSP | Transport | UQCRFS1 |
| **Oxidation-reduction** | | | | |
| 1635227_at[3] | CG10160-RA | ImpL3 | Oxidoreductase | LDHA |
| 1639033_at[3] | CG9432-RB | l(2)01289 | Oxidoreductase | |
| 1638053_at[3] | CG10842-RA | Cyp4p1 | Oxidoreductase | |
| 1637063_at | CG33099-RA | CG33099 | Oxidoreductase | |
| 1623971_at[3] | CG9150-RA | CG9150 | Oxidoreductase | |
| 1623787_at | CG7144-RA | CG7144 | Oxidoreductase | AASS |
| 1627525_a_at | CG1333-RA | Ero1L | Oxidoreductase | ERO1LB |
| 1630885_at[3] | CG12534-RA | Alr | Oxidoreductase | GFER |
| 1627945_at[3] | CG4205-RA | Fdxh | Oxidoreductase | FDX1L |
| 1638006_at | CG10211-RA | CG10211 | Oxidoreductase | |
| 1633687_at | CG13611-RA | CG13611 | Oxidoreductase | |
| 1636759_at | CG8303-RA | CG8303 | Oxidoreductase | |
| 1624571_s_at[3] | CG32857-RA | CG32857 | Oxidoreductase | |
| 1624003_at | CG10639-RA | CG10639 | Oxidoreductase | |
| 1634019_at | CG2064-RA | CG2064 | Oxidoreductase | RDH12 |
| 1640566_at | CG1944-RA | Cyp4p2 | Oxidoreductase | |
| 1632676_s_at[3] | CG11897-RA | CG11897 | Oxidoreductase | |
| 1631856_a_at | CG7361-RB | RFeSP | Oxidoreductase | UQCRFS1 |
| 1633238_at | CG17533-RA | GstEB | Oxidoreductase | |
| 1630258_at | CG4181-RA | GstD2 | Oxidoreductase | |
| **Mitochondrial protein** | | | | |
| 1634658_a_at | CG8772-RD | Nemy | Mitochondrion | GLS |
| 1641606_s_at[3] | CG6608-RB | Tpc1 | Mitochondrion | |
| 1629040_at | CG3476-RA | CG3476 | Mitochondrion | |
| 1633039_at[3] | CG5646-RA | CG5646 | Mitochondrion | |
| 1630885_at[3] | CG12534-RA | Air | Mitochondrion | GFER |
| 1639676_at[3] | CG15173-RA | Ttc19 | Mitochondrion | |
| 1627945_at[3] | CG4205-RA | Fdxh | Mitochondrion | FDX1L |
| 1641511_at | CG7394-RA | TIM14 | Mitochondrion | DNAJC19 |
| 1626460_at | C G2658-RA | CG2658 | Mitochondrion | |
| 1627034_a_a | CG9410-RB | Coq10 | Mitochondrion | |
| 1640220_a_at | CG11779-RA | CG11779 | Mitochondrion | TIMM44 |
| 1627939_a_at[3] | CG2098-RA | Fech | Mitochondrion | FECH |
| 1625763_at | CG2789-RA | CG2789 | Mitochondrion | TSPO |
| 1640566_at | CG1944-RA | Cyp4p2 | Mitochondrion | |
| 1631856_a_at | CG7361-RB | RFeSP | Mitochondrion | UQCRFS1 |
| 1634899_a_at | CG6512-RA | CG6512 | Mitochondrion | AFG3L2 |

[1] GO molecular function/process from www.flybase.org, www.uniprot.org and www.david.abcc.ncifcrf.gov

[2] Human homolog from www.flight.icr.ac.uk

[3] Up-regulated in female and male

**Table S3. Selected genes down-regulated by GMR-Gal4 / UAS-hINS$^{C96Y}$ in male eye imaginal discs**

| Probe set | Transcript | Name | Description (GO)[1] | Homolog[2] |
|---|---|---|---|---|
| **Regulation of transcription** | | | | |
| 1624663_a_at[3] | CG8821-RA | Vismay | DNA binding | |
| 1635500_a_at | CG17228-RA | Prospero | DNA binding | |
| 1631408_at | CG18024-RA | SoxNeuro | DNA binding | SOX1 |
| 1633592_a_at | CG5413-RB | CREG | Transcription repressor | CREG1 |
| 1630105_at | CG3891-RA | Nuclear factor Y-box A | DNA binding | NFYA |
| 1636931_at | CG11491-RD | Broad | DNA binding | QRFPR |
| 1626045_at[3] | CG31318-RA | Rpb4 | Transcription factor | TADA2L |
| 1639732_s_at[4] | CG4881-RA | Spalt-related | RNA polymerase II Transcription factor | SALL1 |
| | | | | |
| **Ion Binding** | | | | |
| 1638311_at | CG12817-RA | CG12817 | Ion binding | |
| 1633488_at[4] | CG3705-RA | astray | Ion binding | PSPH |
| 1630163_at[3] | CG32373-RA | CG32373 | Ion binding | |
| 1629347_at | CG12296-RA | klu | Ion binding | |
| 1633000_a_at | CG7100-RA | CadN | Ion binding | SQRDL |
| 1641652_a_at | CG33166-RB | stet | Ion binding | RHBDL3 |
| 1630504_at | CG13830-RA | CG13830 | Ion binding | |
| 1636602_at | CG11253-RA | CG11253 | Ion binding | |
| 1624001_at | H DC07119 | Scribbler | Ion binding | |
| 1624617_at | CG1665-RA | CG1665 | Ion binding | |
| 1630434_a_at | CG31064-RA | CG31064 | Ion binding | RUFY2 |
| 1639969_at | CG6969-RA | Cardinal | Ion binding | |
| 1635447_at | CG4827-RA | veil | Ion binding | NT5E |
| 1636931_at | CG11491-RD | Broad | Ion binding | QRFPR |
| 1626045_at[3] | CG31318-RA | Rpb4 | Ion binding | TADA2L |
| 1635580_at | CG7037-RB | Cbl | Ion binding | CBL |
| 1624039_at | CG10147-RA | CG10147 | Ion binding | |
| 1638682_a_at | CG11988-RC | Neur | Ion binding | NEURL1B |
| 1640696_at | CG17803-RA | CG17803 | Ion binding | |
| **Enzyme activity** | | | | |
| 1633488_at[4] | CG3705-RA | Astray | Phosphatase | PSPH |
| 1624505_at | CG6113-RA | Lip4 | Lipase | LIPA |
| 1634351_at[4] | CG7860-RA | CG7860 | Asparaginase | |
| 1627360_at | CG31349-RF | Polychaetoid | Guanylate kinase | TJP1 |
| 1626045_at[3] | CG31318-RA | Rpb4 | Acetyltransferase | TADA2L |
| 1623299_at[3] | CG1794-RA | Mmp2 | Endopeptidase | MMP2 |
| 1628149_a_at[3] | CG14895-RB | Pak3 | Protein kinase | PAK3 |

[1] GO molecular function/process from www.flybase.org, www.uniprot.org and www.david.abcc.ncifcrf.gov

[2] Human homolog from www.flight.icr.ac.uk

[3] Down-regulated in female and male

[4] Down-regulated only in female

**Table S4. Functional annotation clustering**

| Database | Keywords | Count | P-value |
|---|---|---|---|
| **Cluster_1** | **Enrichment_Score: 2.4** | | |
| SP_PIR_KEYWORDS | Membrane | 40 | 3.10E-04 |
| GOTERM_CC_FAT | Integral to membrane | 40 | 0.004 |
| GOTERM_CC_FAT | Intrinsic to membrane | 40 | 0.006 |
| SP_PIR_KEYWORDS | Transmembrane | 33 | 0.006 |
| UP_SEQ_FEATURE | Transmembrane region | 19 | 0.026 |
| | | | |
| **Cluster_2** | **Enrichment_Score: 1.38** | | |
| GOTERM_CC_FAT | Cell projection | 7 | 0.021 |
| GOTERM_CC_FAT | Neuron projection | 4 | 0.054 |
| GOTERM_CC_FAT | Dendrite | 3 | 0.063 |
| | | | |
| **Cluster_3** | **Enrichment_Score: 1.32** | | |
| GOTERM_MF_FAT | Iron-sulfur cluster binding | 6 | 0.004 |
| GOTERM_MF_FAT | Metal cluster binding | 6 | 0.004 |
| SP_PIR_KEYWORDS | 2Fe-2S | 3 | 0.022 |
| GOTERM_MF_FAT | 2 iron, 2 sulfur cluster binding | 3 | 0.046 |
| SP_PIR_KEYWORDS | Iron-sulfur | 3 | 0.089 |
| SP_PIR_KEYWORDS | Mitochondrion | 7 | 0.150 |
| SP_PIR_KEYWORDS | Transit peptide | 4 | 0.380 |
| UP_SEQ_FEATURE | Transit peptide:Mitochondrion | 4 | 0.400 |
| | | | |
| **Cluster_4** | **Enrichment_Score: 1.12** | | |
| SMART | DnaJ | 4 | 0.054 |
| INTERPRO | Heat shock protein DnaJ, N-terminal | 4 | 0.059 |
| GOTERM_MF_FAT | Heat shock protein binding | 4 | 0.061 |
| INTERPRO | Molecular chaperone, heat shock protein, Hsp40, DnaJ | 3 | 0.180 |

In total 385 probe sets identified as differentially expressed in either sexes were used in this analysis. 30 of the 385 probe_sets were removed in the most recent Affymetrix annotation, leaving 355. Functional annotation clustering were done using the DAVID web service (ref), with the default choice of annotation terms (excluding GOTERM_BP_FAT) and medium clustering criteria. Changing either the terms or the clustering criteria won't affect the general result. For example, the heat shock protein cluster is always among the top clusters in various settings.

**Table S5 Quantitative RT-PCR**

| Gene | GMR-Gal4 | GMR-Gal4 / UAS-hINS$^{WT}$ | | GMR-Gal4 / UAS-hINS$^{C96Y}$ | |
|------|----------|---------|---------|---------|---------|
| | | WT-6 | WT-24 | M-101 | M-1 |
| Ire1 | 1.05 ± 0.23 | 1.12 ± 0.12 | 1.37 ± 0.28 | 1.99 ± 0.22 (P=0.04) | 1.78 ± 0.23 (P=0.08) |
| PEK[1] | 1.00 ± 0.04 | 0.89 ± 0.11 | 1.11 ± 0.13 | 1.99 ± 0.15 (P=0.003) | 1.98 ± 0.15 (P=0.0007) |
| Hsc70-3[2] | 1.11 ± 0.36 | 0.91 ± 0.27 | 0.74 ±0.07 (P=0.38) | 1.50 ± 0.06 | 2.75 ± 0.22 (P=0.0176) |
| Xbp1 | 1.05 ± 0.25 | 1.50 ± 0.24 | 1.55 ± 0.23 | 2.03 ± 0.17 (P=0.03) | 2.22 ± 0.25 (P=0.03) |

Comparison of gene expression of upstream regulators of UPR in male 3$^{rd}$ instar larval eye imaginal discs is shown. The results was normalized to the expression level of *rp49* and compared to GMR-Gal4 using an unpaired t-*test.* The data are shown as mean ± SE and the exact P-values are shown.

[1] Human homolog of PKR-like ER kinase (PERK)

[2] Human homolog of BiP (also known as GRP78)

**Table S6 Comparison of gene expression levels by microarray and quantitative RT-PCR**

**A. Microarray**

| Gene | GMR-Gal4 | GMR-Gal4 / UAS-hINS$^{WT}$ | | GMR-Gal4 / UAS-hINS$^{C96Y}$ | |
|------|----------|------|------|------|------|
| | | WT-6 | WT-24 | M-101 | M-1 |
| CG3966 | 1.00 ± 0.23 | 0.89 ± 0.00 | 0.87 ± 0.12 | 1.32 ± 0.09 | 6.27 ± 1.08* |
| CG7130 | 1.00 ± 0.02 | 0.96 ± 0.02 | 1.12 ± 0.14 | 1.78 ± 0.04* | 3.90 ± 0.36* |
| CG10420 | 1.00 ± 0.09 | 1.03 ± 0.05 | 0.96 ± 0.07 | 1.98 ± 0.07* | 2.55 ± 0.37* |
| CG10160 | 1.00 ± 0.13 | 1.36 ± 0.15 | 1.03 ±0.15 | 12.63 ± 0.16* | 15.45 ± 1.76* |
| CG9150 | 1.00 ± 0.02 | 1.03 ± 0.02 | 0.85 ± 0.15 | 1.80 ± 0.06* | 2.70 ± 0.56* |

**B. Quantitative RT-PCR**

| Gene | GMR-Gal4 | GMR-Gal4 / UAS-hINS$^{WT}$ | | GMR-Gal4 / UAS-hINS$^{C96Y}$ | |
|------|----------|------|------|------|------|
| | | WT-6 | WT-24 | M-101 | M-1 |
| CG3966 | 1.01 ± 0.08 | 0.98 ± 0.08 | 1.16 ± 0.05 | 1 .97 ± 0.14* | 9.1 ± 0.93* |
| CG7130 | 1.00 ± 0.02 | 0.84 ± 0.21 | 0.99 ± 0.03 | 1.66 ± 0.13* | 4.03 ± 0.71* |
| CG10420 | 1.01 ± 0.08 | 1.12 ± 0.06 | 1.03 ± 0.14 | 4.81 ± 0.43* | 11.77 ± 0.44* |
| CG10160 | 1.01 ± 0.09 | 1.11 ± 0.35 | 0.70 ± 0.22 | 234.64 ± 80.76* | 378 ± 64.08* |
| CG9150 | 1.02 ± 0.12 | 1.18 ± 0.32 | 1.02 ± 0.08 | 2.49 ± 0.07* | 3.73 ± 0.38* |

Gene expression was normalized to the expression level of *rp49* and compared to GMR-Gal4 using an unpaired t-*test*. Data are shown as mean ± SE.

* P < 0.05

**Table S7 Number of Missing Bristles**

| | ap>>hINS[C96Y]/ CyO | ap>>hINS[C96Y] X W1118 | Sco X w1118 |
|---|---|---|---|
| Female | 6.56 ± 1.80 | 8.32 ± 2.14 | 9.20 ±1.44 |
| Male | 10.92 ± 2.33 | 13.36 ± 2.48 | 9.36 ±1.32 |

We recombined the ap-GAL4 driver onto the chromosome containing the M-1 UAS-hINS[C96Y] transgene to create ap>>hINS[C96Y]. Sco (Scutoid) also reduces bristles on the notum to approximately the same level as mutant hINS. We used Sco in a control cross to investigate whether it has a sex-biased bristle phenotype. The fact that it does not suggests that the sex-biased phenotype produced by mutant hINS may not be through sex-specific inputs to bristle formation but rather through a physiological difference in the response of male and female somatic cells to mutant hINS protein. Data are shown as mean ± SD.

**Table S8 Missing or displaced companiform sensilla**

| Sex | en>>hINS[WT] | en>>hINS[C96Y] | dpp>>hINS[WT] | dpp>>hINS[C96Y] |
|---|---|---|---|---|
| Female | 0 | 7 | 0 | 11 |
| Male | 0 | 4 | 1 | 14 |

We examined four sensilla — one on the ACV and three along L3 (see Figure S5) — from one wing in 10 individuals. Tabulated are the total number of displaced sensillae out of 40.

**Table S9 Correlation between mutations and sexes**

| | Drop | Lobe | GMR>>hINS[C96Y] | ap>>hINS[C96Y] |
|---|---|---|---|---|
| Drop | 0.63* | 0.2 | 0.06 | -0.19 |
| Lobe | 0.09 | 0.53* | 0.08 | -0.12 |
| GMR>>hINS[C96Y] | 0.13 | -0.07 | 0.88* | -0.23 |
| ap>>hINS[C96Y] | -0.19 | 0.29 | -0.17 | 0.67* |

Diagonal elements are correlations between the two sexes for each mutation group; the upper triangle contains correlations between males in the respective mutation groups, and numbers in the lower triangle area are for females. Pearson's correlation is calculated for all pairs except those involving the bristle number, in which case the Spearman's correlation (rank correlation) is calculated. * indicates a correlation test $P < 0.05$ (P-values corrected for multiple testing using Bonferroni method)

**Table S10 Effect of genetic variation on eye area and bristle number in females**

| | GMR>>INS$^{C96Y}$ | Drop | Lobe | ap>>INS$^{C96Y}$ |
|---|---|---|---|---|
| **DGRP Line** | **Eye area** | **Eye area** | **Eye area** | **Number of Bristles** |
| RAL-208 | 64.58 ± 6.99 | 14.60 ± 4.01 | 94.66 ± 5.69 | 23.1 ± 2.0 |
| RAL-301 | 66.59 ± 5.81 | 16.88 ± 3.88 | 97.50 ± 4.73 | 25.2 ± 1.0 |
| RAL-303 | 61.25 ± 6.32 | 14.01 ± 1.88 | 94.54 ± 4.86 | 24.5 ± 1.5 |
| RAL-304 | 66.87 ± 27.85 | 14.63 ± 2.43 | 97.36 ± 5.79 | 23.8 ± 1.9 |
| RAL-307 | 80.10 ± 8.11 | 15.08 ± 4.20 | 101.84 ± 6.73 | 25.7 ± 0.5 |
| RAL-313 | 68.62 ± 5.47 | 14.22 ± 4.33 | 91.99 ± 8.85 | 25.8 ± 0.5 |
| RAL-315 | 47.58 ± 4.55 | 9.61 ± 1.82 | 82.04 ± 4.74 | 23.9 ± 2.2 |
| RAL-324 | 67.72 ± 4.65 | 11.02 ± 1.92 | NA | 24.9 ± 0.9 |
| RAL-335 | 63.11 ± 6.28 | 12.38 ± 1.82 | 96.48 ± 7.63 | 18.6 ± 2.1 |
| RAL-357 | 68.49 ± 11.45 | 13.53 ± 2.19 | 86.01 ± 9.87 | 25.6 ± 0.6 |
| RAL-358 | 70.18 ± 4.91 | 16.39 ± 3.63 | 95.36 ± 6.01 | 25.2 ± 1.3 |
| RAL-360 | 69.02 ± 6.85 | 13.30 ± 2.60 | 95.21 ±13.73 | 23.9 ± 1.4 |
| RAL-362 | 37.29 ± 4.37 | 24.21 ± 8.91 | 98.77 ± 7.44 | 24.8 ± 1.0 |
| RAL-375 | 55.85 ± 6.87 | 14.05 ± 1.83 | 91.55 ± 5.53 | 24.0 ± 1.3 |
| RAL-379 | 59.42 ± 6.15 | 16.02 ± 3.10 | 98.02 ± 5.42 | 24.4 ± 1.2 |
| RAL-380 | 69.96 ± 4.07 | 11.89 ± 1.83 | 93.38 ± 5.33 | 20.5 ± 2.1 |
| RAL-391 | 48.63 ± 6.85 | 15.43 ± 4.56 | 87.50 ± 8.92 | 23.8 ± 2.2 |
| RAL-399 | 43.31 ± 8.51 | 13.75 ± 2.93 | 97.43 ± 8.47 | 25.2 ± 0.9 |
| RAL-427 | 53.88 ± 12.51 | 19.65 ± 6.10 | 96.57 ± 9.69 | 25.8 ± 0.5 |
| RAL-437 | 70.21 ± 8.30 | 16.73 ± 3.03 | 101.10 ± 5.44 | 23.5 ± 1.4 |
| RAL-486 | 60.40 ± 5.79 | 17.66 ± 3.51 | 96.22 ± 6.41 | 23.4 ± 1.3 |
| RAL-426 | 63.41 ± 8.99 | 12.88 ± 1.58 | 95.91 ± 7.65 | 23.3 ± 1.5 |
| RAL-517 | 71.37 ± 5.08 | 10.40 ± 1.87 | 104.17 ± 6.48 | 24.6 ± 1.2 |
| RAL-555 | 56.32 ± 8.91 | 13.77 ± 3.18 | 101.21 ± 5.35 | 23.0 ± 1.9 |
| RAL-639 | 28.56 ± 15.68 | 13.60 ± 2.78 | 95.42 ± 5.39 | 23.2 ± 2.1 |
| RAL-707 | 71.94 ± 6.69 | 14.27 ± 2.12 | 103.78 ± 6.82 | 25.2 ± 0.9 |
| RAL-712 | 59.56 ± 6.67 | 13.57 ± 3.86 | 73.41 ± 22.28 | 25.7 ± 0.7 |
| RAL-730 | 61.14 ± 10.99 | 14.54 ± 2.89 | 101.73 ± 5.57 | 25.3 ± 1.0 |
| RAL-732 | 55.86 ± 7.85 | 12.50 ± 1.80 | 96.72 ± 8.16 | 23.9 ± 1.5 |
| RAL-765 | 50.59 ± 11.68 | 13.56 ± 1.74 | 62.45 ± 20.45 | 24.0 ± 0.8 |
| RAL-774 | 67.71 ± 7.20 | 13.68 ± 2.19 | 91.80 ± 7.02 | 25.8 ± 0.4 |
| RAL-786 | 54.59 ± 7.48 | 12.54 ± 1.76 | 94.26 ± 6.55 | 25.6 ± 0.6 |
| RAL-799 | 59.45 ± 5.81 | 13.21 ± 3.85 | 99.35 ± 7.31 | 24.0 ± 1.2 |
| RAL-820 | 55.22 ± 10.40 | 14.10 ± 2.23 | 101.15 ± 5.98 | 25.1 ± 1.0 |
| RAL-852 | 72.05 ± 6.21 | 14.60 ± 3.26 | 81.69 ± 16.53 | 24.8 ± 0.9 |
| RAL-365 | 76.59 ± 4.88 | 14.03 ± 3.49 | 99.59 ± 7.13 | 25.4 ± 0.9 |
| RAL-705 | 52.63 ± 7.89 | 13.68 ± 2.14 | 95.56 ± 7.65 | 24.4 ± 1.1 |
| RAL-714 | 68.74 ± 6.30 | 12.54 ± 3.62 | NA | 25.0 ± 1.2 |

The eye area or dorsal bristle number were measured in 10 individuals. The eye area (pixels $\times 10^3$) and number of bristles are shown as mean ± SE.

**Table S11 Effect of genetic variation on eye area and bristle number in males**

|  | GMR>>INS$^{C96Y}$ | Drop | Lobe | ap>>INS$^{C96Y}$ |
|---|---|---|---|---|
| DGRP Line | Eye area | Eye area | Eye area | Number of Bristles |
| RAL-208 | 25.55 ± 4.76 | 9.15 ± 2.71 | 80.79 ± 7.57 | 19.8 ± 1.5 |
| RAL-301 | 32.33 ± 5.56 | 11.16 ± 3.14 | 73.46 ± 6.20 | 18.3 ± 3.9 |
| RAL-303 | 31.41 ± 6.53 | 9.87 ± 2.99 | 73.67 ± 5.44 | 23.3 ± 1.6 |
| RAL-304 | 39.92 ± 7.34 | 10.87 ± 4.85 | 76.08 ± 6.08 | 21.0 ± 2.0 |
| RAL-307 | 46.12 ± 7.40 | 15.56 ± 4.26 | 91.04 ± 7.14 | 23.4 ± 1.5 |
| RAL-313 | 36.76 ± 6.94 | 8.25 ± 1.89 | 86.30 ± 5.75 | 24.7 ± 1.4 |
| RAL-315 | 22.38 ± 2.72 | 6.68 ± 2.12 | 73.13 ± 4.60 | 19.8 ± 1.8 |
| RAL-324 | 44.29 ± 4.23 | 7.98 ± 2.92 | 75.63 ± 4.38 | 23.4 ± 1.1 |
| RAL-335 | 35.77 ± 4.32 | 8.02 ± 2.39 | 77.02 ± 5.83 | 11.0 ± 1.1 |
| RAL-357 | 36.14 ± 4.49 | 11.80 ± 3.11 | 81.68 ± 6.35 | 25.1 ± 1.2 |
| RAL-358 | 34.04 ± 6.96 | 11.84 ± 4.18 | 82.04 ± 7.65 | 22.4 ± 2.0 |
| RAL-360 | 30.18 ± 10.04 | 12.16 ± 3.32 | 85.21 ± 6.17 | 22.4 ± 1.8 |
| RAL-362 | 70.52 ± 7.21 | 13.10 ± 5.01 | 82.63 ± 4.26 | 24.0 ± 1.1 |
| RAL-375 | 29.00 ± 5.43 | 12.65 ± 2.79 | 83.53 ± 6.19 | 20.1 ± 2.5 |
| RAL-379 | 27.05 ± 5.47 | 12.22 ± 3.35 | 85.32 ± 4.69 | 21.6 ± 2.2 |
| RAL-380 | 36.80 ± 6.07 | 9.01 ± 2.70 | 80.68 ± 6.72 | 13.7 ± 2.2 |
| RAL-391 | 13.83 ± 3.61 | 13.57 ± 4.38 | 70.53 ± 6.69 | 21.9 ± 1.9 |
| RAL-399 | 10.65 ± 3.48 | 10.03 ± 3.37 | 81.53 ± 5.45 | 17.9 ± 2.5 |
| RAL-427 | 25.96 ± 5.16 | 14.30 ± 3.63 | 87.13 ± 4.94 | 25.6 ± 0.7 |
| RAL-437 | 33.82 ± 7.29 | 13.15 ± 3.60 | 85.73 ± 5.73 | 20.7 ± 2.3 |
| RAL-486 | 30.94 ± 6.10 | 11.30 ± 4.44 | 89.08 ± 5.77 | 22.4 ± 1.8 |
| RAL-426 | 30.93 ± 6.49 | 9.16 ± 2.82 | 80.82 ± 5.12 | 20.6 ± 2.0 |
| RAL-517 | 29.32 ± 4.64 | 5.27 ± 1.09 | 90.12 ± 3.94 | 21.5 ± 1.9 |
| RAL-555 | 23.27 ± 4.74 | 9.55 ± 2.63 | 83.83 ± 5.54 | 21.9 ± 1.9 |
| RAL-639 | 7.13 ± 4.47 | 8.40 ± 2.25 | 84.64 ± 6.59 | 21.1 ± 2.2 |
| RAL-707 | 34.95 ± 7.72 | 9.97 ± 3.12 | 78.70 ± 7.26 | 23.4 ± 1.4 |
| RAL-712 | 31.96 ± 7.29 | 9.55 ± 2.42 | 80.80 ± 6.80 | 23.8 ± 1.4 |
| RAL-730 | 23.92 ± 9.50 | 15.32 ± 7.03 | 77.42 ± 5.74 | 22.6 ± 1.5 |
| RAL-732 | 21.65 ± 5.90 | 10.90 ± 4.01 | 84.35 ± 6.96 | 19.6 ± 2.5 |
| RAL-765 | 13.46 ± 5.90 | 9.62 ± 2.12 | 80.69 ± 5.55 | 19.7 ± 1.6 |
| RAL-774 | 24.75 ± 5.48 | 8.49 ± 1.94 | 69.95 ± 6.60 | 25.5 ± 0.7 |
| RAL-786 | 22.51 ± 5.73 | 11.58 ± 3.17 | 83.99 ± 6.38 | 24.7 ± 1.5 |

| | | | |
|---|---|---|---|
| RAL-799 | 24.51 ± 7.25 | 9.10 ± 2.72 | 81.61 ± 3.29 | 21.0 ± 2.1 |
| RAL-820 | 21.97 ± 6.59 | 10.59 ± 4.59 | 82.90 ± 4.00 | 24.3 ± 1.7 |
| RAL-852 | 34.65 ± 7.39 | 10.82 ± 2.65 | 86.58 ± 4.90 | 14.1 ± 2.8 |
| RAL-365 | 43.91 ± 5.69 | 9.45 ± 2.96 | 77.41 ± 4.70 | 23.9 ± 1.4 |
| RAL-705 | 13.90 ± 3.58 | 10.40 ± 4.10 | 82.48 ± 7.07 | 18.8 ± 2.4 |
| RAL-714 | 42.56 ± 5.22 | 10.05 ± 3.41 | 84.19 ± 4.99 | 24.0 ± 1.9 |

The eye area or dorsal bristle number were measured in 10 individuals. The eye area (pixels $\times 10^3$) and number of bristles are shown as mean ± SE.